\documentclass[longauth]{aa}
\usepackage{placeins}

\usepackage{graphicx}
\usepackage[flushleft]{threeparttable}

\usepackage{txfonts}
\usepackage[citecolor=blue, linkcolor=blue]{hyperref}
\hypersetup{colorlinks=True}
\newcommand{\Msun}{M$_{\odot}$}
\newcommand{\Mstar}{M$_{\star}$\,}
\newcommand{\xiion}{$\xi_{\rm ion}$\,}
\newcommand{\halpha}{H${\alpha}$\,}
\newcommand{\hbeta}{H${\beta}$\,}

\begin{document} 

\title{The ionizing photon production efficiency of star-forming galaxies at $z\sim 4-10$}
 
\author{Llerena, M.\inst{1}\fnmsep\thanks{\email{mario.llerenaona@inaf.it}}
\and Pentericci, L.\inst{1}
\and Napolitano, L. \inst{1,2}
\and Mascia, S. \inst{1,3}
\and Amor\'{i}n, R. \inst{4} 
\and Calabr{\`o}, A.\inst{1}
\and Castellano, M.\inst{1}
\and Cleri, N. J. \inst{5,6,7}
\and Giavalisco, M. \inst{8}
\and Grogin, N. A. \inst{9}
\and Hathi, N. P. \inst{9}
\and Hirschmann, M. \inst{10}
\and Koekemoer, A. M.\inst{9}
\and Nanayakkara, T. \inst{11} 
\and Pacucci, F. \inst{12,13}
\and Shen, L. \inst{14}
\and Wilkins, S. M.  \inst{15,16}
\and Yoon, I. \inst{17}
\and Yung, L.~Y.~A.~ \inst{9}
\and Bhatawdekar, R. \inst{18}
\and Lucas, R. A. \inst{9}
\and Wang, X. \inst{19,20,21}
\and Arrabal Haro, P. \inst{22}\fnmsep\thanks{{NASA Postdoctoral Fellow}}
\and Bagley, M. B. \inst{23} 
\and Finkelstein, S. L. \inst{23}
\and Kartaltepe, J. S. \inst{24}
\and Merlin, E. \inst{1}
\and Papovich, C. \inst{14,25}
\and Pirzkal, N. \inst{26}
\and Santini, P. \inst{1}
          }

   \institute{INAF - Osservatorio Astronomico di Roma, Via di Frascati 33, 00078, Monte Porzio Catone, Italy
   \and Dipartimento di Fisica, Università di Roma Sapienza, Città Universitaria di Roma - Sapienza, Piazzale Aldo Moro, 2, 00185, Roma, Italy
   \and Institute of Science and Technology Austria (ISTA), Am Campus 1, A-3400 Klosterneuburg, Austria
   \and Instituto de Astrof\'{i}sica de Andaluc\'{i}a (CSIC), Apartado 3004, 18080 Granada, Spain
   \and {Department of Astronomy and Astrophysics, The Pennsylvania State University, University Park, PA 16802, USA}
\and {Institute for Computational and Data Sciences, The Pennsylvania State University, University Park, PA 16802, USA}
\and {Institute for Gravitation and the Cosmos, The Pennsylvania State University, University Park, PA 16802, USA}
\and University of Massachusetts Amherst, 710 North Pleasant Street, Amherst, MA 01003-9305, USA
\and Space Telescope Science Institute, 3700 San Martin Drive, Baltimore, MD 21218, USA
\and{Institute of Physics, Laboratory of Galaxy Evolution, Ecole Polytechnique Fédérale de Lausanne (EPFL), Observatoire de Sauverny, 1290 Versoix, Switzerland}
\and{Centre for Astrophysics and Supercomputing, Swinburne University of Technology, PO Box 218, Hawthorn, VIC 3122, Australia}
\and Center for Astrophysics $\vert$ Harvard \& Smithsonian, Cambridge, MA 02138, USA
   \and Black Hole Initiative, Harvard University, Cambridge, MA 02138, USA
   \and{Department of Physics and Astronomy, Texas A\&M University, College Station, TX, 77843-4242 USA}
   \and{Astronomy Centre, University of Sussex, Falmer, Brighton BN1 9QH, UK}
 \and{Institute of Space Sciences and Astronomy, University of Malta, Msida MSD 2080, Malta}
 \and{National Radio Astronomy Observatory, 520 Edgemont Road, Charlottesville, VA 22903, USA}
   \and European Space Agency (ESA), European Space Astronomy Centre (ESAC), Camino Bajo del Castillo s/n, 28692 Villanueva de la Cañada, Madrid, Spain
   \and {School of Astronomy and Space Science, University of Chinese Academy of Sciences (UCAS), Beijing 100049, China}
\and {National Astronomical Observatories, Chinese Academy of Sciences, Beijing 100101, China}
\and{Institute for Frontiers in Astronomy and Astrophysics, Beijing Normal University, Beijing 102206, China} 
   \and Astrophysics Science Division, NASA Goddard Space Flight Center, 8800 Greenbelt Rd, Greenbelt, MD 20771, USA
   \and Department of Astronomy, The University of Texas at Austin, Austin, TX, USA
   \and Laboratory for Multiwavelength Astrophysics, School of Physics and Astronomy, Rochester Institute of Technology, 84 Lomb Memorial Drive, Rochester, NY 14623, USA
   \and George P. and Cynthia Woods Mitchell Institute for Fundamental Physics and Astronomy, Department of Physics and Astronomy, Texas A\&M University, College Station, TX, USA
   \and ESA/AURA Space Telescope Science Institute, Baltimore, USA
    } 

   \date{Received ; accepted }

%

%
%
%

%
%

%

%
%
%
%
%
%
%
%
%

 
  \abstract
   {Investigating the ionizing emission of star-forming galaxies {and the escape fraction of ionizing photons} is critical to understanding their contribution to reionization and their impact on the surrounding environment. The number of ionizing photons available to reionize the intergalactic medium (IGM) depends not only on the abundance of galaxies but also on their efficiency in producing ionizing photons (\xiion). This quantity is thus fundamental to quantify the role of faint versus bright sources in driving this process, as we must assess their relative contribution to the total ionizing emissivity.
   }   
   {Our goal is to estimate the \xiion using Balmer lines (H$\alpha$ or H$\beta$) in a sample of {761} galaxies at $4\leq z \leq 10$ selected from different JWST spectroscopic surveys. We aim to determine the redshift evolution of \xiion and the relation of \xiion with the physical properties of the galaxies.
   }
   {We used the available HST and JWST photometry to perform a spectral energy distribution (SED) fitting in the sample to determine their physical properties and relate them with \xiion. We used the BAGPIPES code for the SED fitting and assumed a delayed exponential model for the star formation history. We used the NIRSpec spectra from prism or grating configurations to estimate Balmer luminosities and then constrained \xiion values after dust correction.
   }
   {We find a mean value of 10$^{25.22}$Hz erg$^{-1}$ for \xiion in the sample with an observed scatter of 0.42dex. We find an increase of the median values of \xiion with redshift from {10$^{25.09}$Hz erg$^{-1}$ at $z\sim4.18$ to 10$^{25.28}$Hz erg$^{-1}$ at $z\sim7.14$}, which confirmed the redshift evolution of \xiion found in other works. Regarding the relation with physical properties, we find a decrease of \xiion with increasing stellar mass, indicating that low-mass galaxies are efficient producers of ionizing photons. We also find an increase of \xiion with increasing specific star formation rate (sSFR) and increasing UV absolute magnitude, which indicates that faint galaxies and with high sSFR are also efficient producers. We also investigated the relation of \xiion with the rest-frame equivalent with (EW) of [OIII]$\lambda$5007 and find that galaxies with the higher EW([OIII]$\lambda$5007) are the more efficient producers of ionizing photons with the best-fit leading to the relation log(\xiion)=0.43$\times\log$(EW[OIII])+{23.99}. Similarly, we find that galaxies with the higher O32=[OIII]$\lambda$5007/[OII]$\lambda\lambda$3727,3729 ratios and lower gas-phase metallicities (based on the R23=($[O III]\lambda\lambda$4959,5007+$[O II]\lambda\lambda$3727,3729)/H$\beta$ calibration) show higher \xiion values.
   }

\titlerunning{The ionizing photon production efficiency of star-forming galaxies at $z\sim 4-10$}
\authorrunning{Llerena, M. et al.}

\maketitle
%

\section{Introduction} \label{sec:intro}

The Epoch of Reionization (EoR) marks a critical phase in the evolution of the Universe, during which the intergalactic medium (IGM) became transparent to Lyman Continuum (LyC) radiation (energy $\geq$ 13.6 eV). Observations suggest that this epoch concluded around redshift $z\sim 6$ \citep[][]{Fan2006,Yang2020}, though some research indicates reionization may have extended closer to $z \sim 5$ \citep{Bosman2022}. The consensus is that young, massive stars within galaxies played a major role in this transformation by producing large quantities of LyC photons, which escaped the interstellar medium (ISM) and ultimately ionized the IGM \citep[e.g.,][]{Rosdahl2018}. However, debate persists over whether faint, low-mass galaxies, bright, massive galaxies, or a combination of both contributed the most to the reionization photon budget \citep{Finkelstein2019,Naidu2020,Robertson2022}. Additionally, the role of active galactic nuclei (AGN) in reionization may be more significant than previously thought, with some studies suggesting AGNs and their host galaxies account for over 10\% of the photon budget \citep{Maiolino2023,Madau2024}.

Investigating the ionizing emission of star-forming galaxies is critical to understanding their contribution to reionization and their impact on the surrounding environment. The number of ionizing photons available to reionize the IGM depends not only on the abundance of galaxies but also on their efficiency in producing LyC radiation and the fraction of this radiation that escapes into the IGM. This quantity is thus fundamental to quantify the role of faint versus bright sources in driving this process, as we must assess their relative contribution to the total ionizing emissivity ($\dot{n}_{ion}$) -- the number of ionizing photons emitted per unit time and comoving volume. This is commonly expressed as:
\begin{equation}
\dot{n}_{ion}=\rho_{UV}\xi_{ion}f_{esc}
\end{equation}
where $\rho_{UV}$ is the comoving total UV luminosity density (erg s$^{-1}$ Hz$^{-1}$ Mpc$^{-3}$
), $f_{esc}$ is the
fraction of LyC photons that escape galaxies to ionize intergalactic hydrogen,
and \xiion is the ionizing photon production efficiency (Hz erg$^{-1}$
) that indicates the number
of LyC photons per unit UV luminosity density the stellar populations in
galaxies generate. The quantity $\dot{n}_{ion}$ represents the comoving density of LyC photons produced per unit time available for ionizing hydrogen in the IGM.
Each of these three components carries significant uncertainties in both modeling and observation, making it extremely difficult to model them consistently. This challenge arises because the underlying physical processes span many orders of magnitude in scale.

For galaxies to be the primary drivers of reionization, relatively high escape fractions of ionizing photons are required, typically ranging from 10\% to 20\% \citep{Robertson2013,Finkelstein2019,Naidu2020,Yung2020}. Another crucial factor is the \xiion. {Standard reionization models \citep[e.g.,][]{Madau1999,Robertson2013} find a value of log\xiion[Hz erg$^{-1}$]=25.2-25.3 for star-forming galaxies which is in agreement with observations at $z\sim 4$ \citep[e.g.,][]{Bouwens2016}. Throughout the paper, we will consider a value of log\xiion[Hz erg$^{-1}$]=25.27 as the canonical value based on observations}. Recent observations, which extend up to redshift \( z \sim 9 \), show that \xiion increases at higher redshifts \citep{Tang2023,Simmonds2024,Pahl2024,Atek2024,Begley2025}. This rise in \xiion implies that lower escape fractions are needed for galaxies to have been the main contributors to reionization. 

The launch of the James Webb Space Telescope \citep[JWST, ][]{Gardner2006,Gardner2023} has provided unprecedented access to the rest-frame optical range at high redshifts, offering valuable new data to improve our understanding of \xiion and its variation across the galaxy population. As a result, studies focused on how $f_{esc}$ and \xiion evolve with galaxy properties, especially at high redshifts, are critical for advancing our understanding of the EoR. A key observation is that stellar mass correlates with its ability to produce ionizing photons \citep{Castellano2023}. 

A detailed and thorough study of \xiion requires large and
representative samples with spectroscopic \halpha and \hbeta
observations to measure nebular line luminosity directly and
properly correct for dust. One method of measuring \xiion that enables us to constrain its value for individual galaxies is by using nebular recombination lines, e.g., \halpha. In an ionization-bounded nebula, the recombination rate balances the rate of photons with energies at or above 13.6 eV that are either emitted from the star or produced during recombination to the hydrogen ground level.

In this paper, we built a sample of {761} spectroscopically confirmed galaxies with NIRSpec \citep{Jakobsen2022} at $4\leq z\leq 10$ from different JWST programs to estimate their individual \xiion values based on Balmer lines. We also estimated their physical properties based on fitting their Spectral Energy Distribution (SED) using the available NIRCam \citep{Beichman2012} photometry of these sources. This paper is organized as follows: in Sec. \ref{sec:sample}, we describe the sample selection. In Sec. \ref{sec:analysis}, we describe the data analysis which includes emission line measurements, SED fitting, UV luminosity estimations, and \xiion calculations for individual galaxies. In Sec. \ref{sec:redshift}, we discuss the evolution of \xiion with redshift. {In Sec. \ref{sec:xi_muv} we show the relation between \xiion and UV absolute magnitude (M$_{UV}$), and also the evolution with redshift. In Sec. \ref{sec:total_emissivity} we analyze the total ionizing emissivity using the relation found in this paper.} In Sec. \ref{sec:properties}, we show the dependency of \xiion with the physical properties which include stellar mass and specific star formation rate (sSFR). In Sec. \ref{sec:EW}, we show the relation of \xiion with the rest-frame equivalent width (EW) of [OIII]$\lambda$5007. In Sec. \ref{sec:metallicity}, we discuss the relation of \xiion with line flux ratios and gas-phase metallicity. Finally, we present our {summary} and conclusions in Sec. \ref{sec:conclusions}.

Throughout this paper, we adopt a $\Lambda$-dominated flat universe with $\Omega_\Lambda = 0.7$, $\Omega_M = 0.3,$ and H$_0= 70$ km s$^{-1}$ Mpc$^{-1}$. All magnitudes are quoted in the AB system \citep{Oke1983}. Equivalent widths are quoted in the rest frame and are positive for emission lines. We consider log(O/H)$_{\odot}=8.69$ \citep{Asplund2009}.

\section{Data and sample selection}\label{sec:sample}
We selected galaxies at redshift $z=4-10$ in order to include the rest-UV and \hbeta in the prism configuration using NIRSpec. We selected galaxies from the Cosmic Evolution Early Release Science survey \citep[CEERS, ][]{Finkelstein2023}, the JWST Advanced Deep Extragalactic Survey \citep[JADES, ][]{Eisenstein2023}, the GLASS survey \citep[][]{Treu2022} and the program GO-3073 (PI: M. Castellano). We describe the selection in each survey in the next subsections. Our final sample includes {790} galaxies.
\begin{figure}[t!]
    \centering
    \includegraphics[width=\columnwidth]{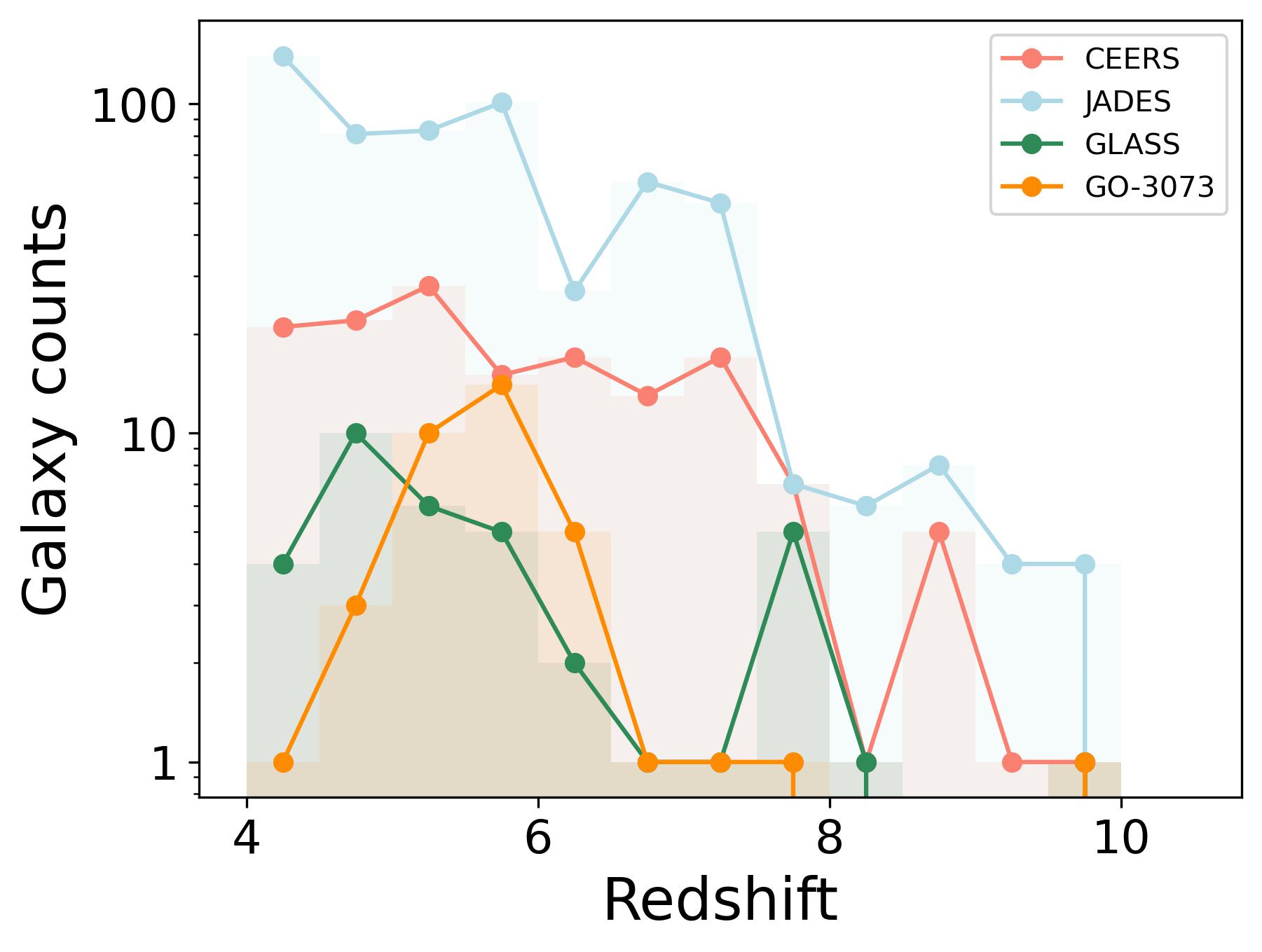}
    \caption{Redshift distribution of the selected sample of spectroscopically confirmed galaxies at $4\leq z\leq 10$ in the different surveys considered in this work as described in the main text.}
    \label{fig:redshift}
\end{figure}
\subsection{The CEERS survey}
The complete CEERS program involved imaging with the NIRCam short and long-wavelength channels in ten pointings, observed as coordinated parallels to primary observations with the NIRSpec and the MidInfrared Instrument \citep[MIRI;][]{Wright2015}. In this paper, we used photometric data from NIRCam and NIRSpec spectra. In total, 10 pointings were taken with NIRCam imaging, including seven filters per pointing (F115W, F150W, F200W, F277W, F356W, F410M, and F444W).

We used version v0.51.2 of the CEERS Photometric Catalogs (Finkelstein et al. \textit{in prep}). The catalog contains 101808 sources. The NIRCam images used are publicly available, and we refer the reader to  \cite{Bagley2022} for a complete description of the data reduction. For the pointings 1, 2, 3, and 6, the images are available in the Data Release 0.5\footnote{\url{https://ceers.github.io/dr05.html}}, while for the pointings 4, 5, 7, 8, 9, and 10, the images are available in the Data Release 0.6  \footnote{\url{https://ceers.github.io/dr06.html}}. 

A complete description of the photometric catalog will be presented in Finkelstein et al. \textit{in prep}. Briefly, the photometry was performed with SExtractor \citep[v2.25.0; ][]{Bertin1996} with F277W and F356W as the detection image. The fiducial fluxes were measured in small Kron apertures corrected by large-scale flux, following the methodology in \cite{Finkelstein2023}. {This photometry also makes use of archival HST data on this field from
programs including CANDELS \citep{Koekemoer2011,Grogin2011}.} 

The CEERS survey also includes six NIRSpec pointings, numbered p4, p5, p7, p8, p9, and p10. Each of these pointings has observations with the three NIRSpec medium resolution gratings (G140M, G235M, and G395M) and with the low-resolution Prism. Two more fields, p11 and p12, were observed with the prism in February 2023 because the prism observations p9 and p10 were severely impacted by a short circuit. The grating set covers from 0.97–5.10$\mu$m with a resolving power $R=\lambda / \Delta\lambda$ of $\sim$1000, while the prism covers from 0.60-5.30$\mu$m with a resolving power of $30 < R < 300$, depending on the wavelength.  

We adopt the NIRSpec data produced by the CEERS collaboration using the STScI JWST Calibration Pipeline\footnote{\url{https://github.com/spacetelescope/jwst}} \citep{Bushouse2022}.
Specifically, we used the JWST pipeline to perform standard reductions, including the removal of dark current and bias, flat-fielding, background, photometry, wavelength, and slitloss correction for each exposure.
We also perform additional reductions to remove the $1/f$ noise and the snowballs.
The 2D spectra of each target were then rectified and combined to generate the final 2D spectra.
The details of the data reduction are presented in \citet[][]{ArrabalHaro2023} and Arrabal Haro et al. \textit{in prep.} 

We selected galaxies with NIRSpec spectra at redshift $4 \leq z\leq 10$. We consider the low limit $z=4$ to ensure that the range $\sim$ 1200\r{A} is observed in the prism configuration and we consider the high redshift limit $z=10$ to ensure H$\beta$ is observed. We do not consider NIRSpec data observed in pointings 9 and 10, since due to a short circuit issue, they are contaminated and lack secure flux calibrations. We selected galaxies with available NIRCam photometry. Our sample consists of {148} galaxies, {140} with prism observations and {32} with MR grating observations. The redshift distribution of the sample is shown in red in Fig. \ref{fig:redshift}.

\subsection{The JADES survey}
We considered the third public data release of the survey \citep{DEugenio2024} to select our sample. JADES provides both imaging and spectroscopy in the two GOODS fields. Spectroscopy consists of medium-depth and deep NIRSpec/MSA spectra of 4000 targets, covering the spectral range 0.6-5.3$\mu$m and observed with both the low-dispersion prism (R=30-300) and the three medium-resolution gratings (R=500-1500). We refer to \citep{DEugenio2024} for a complete description of observations, data reduction, sample selection, and target allocation. A total of 2053 redshifts were measured from multiple emission lines in this data release. The photometric catalog includes observation using the NIRCam wide-broad filters F090W, F115W, F150W, F200W, F277W, F356W, F444W, and the medium-broad filters F182M, F210M, F335M, F410M, F430M, F460M, F480M. 

We selected galaxies with NIRSpec spectra at redshift $4 \leq z\leq 10$. We selected galaxies with NIRCam photometry. Our sample thus includes  314 galaxies {(240 with MR and 314 with prism)} in the GOODS-South field and 255 galaxies {(239 with MR and 246 with prism)} in the GOODS-North field. The redshift distribution of the total sample of 569 galaxies is shown in lightblue in Fig. \ref{fig:redshift}.

\subsection{The GLASS survey}

The GLASS survey obtained deep observations of galaxies in the Hubble Frontier Field cluster, Abell 2744 with both NIRCam photometry and NIRSpec spectroscopy. The public catalog includes galaxies observed with prism and the high-resolution G140H, G235H, G395H grating. The public spectroscopic data release is found in \cite{Mascia2024}. We use the NIRSpec
data reduced by the Cosmic Dawn Center, which is published on the DAWN JWST Archive (DJA)\footnote{\url{https://dawn-cph.github.io/dja/spectroscopy/nirspec/}}. The photometric catalog includes observations using the NIRCam wide-broad filters F115W, F150W, F200W, F277W, F356W, F444W, and the medium-broad filters F410M \citep{Merlin2024}.

We selected galaxies with NIRSpec spectra at redshift $4 \leq z\leq 10$ and with   NIRCam photometry. Our sample includes 36 galaxies {(23 with HR and 25 with prism)} and its redshift distribution of  is shown in green in Fig. \ref{fig:redshift}.

\subsection{The GO-3073 program}

The program observed the cluster Abell-2744 using the prism configuration of NIRSpec. In this paper, we considered data obtained during the first epoch (October 24, 2023), divided into three visits, each with an exposure time of 6567 s. Unfortunately, an electrical short affected the third visit, so we excluded it from the final reduction. The data were processed using version 1.13.4 of the STScI Calibration Pipeline, with Calibration Reference Data System (CRDS) mapping 1197. Further details on observations and data reduction are provided in \cite{Napolitano2024}. The photometric catalog includes observations using the NIRCam wide-broad filters F090W, F115W, F150W, F200W, F277W, F356W, F444W \citep{Merlin2024}.

We selected 37 galaxies with $4<z<10$ with NIRSpec, NIRCam photometry, and secure spectroscopic redshift. The redshift distribution of the sample is shown in orange in Fig. \ref{fig:redshift}.

\section{Data analysis}\label{sec:analysis}
\subsection{AGN removal and complete subsample}\label{sec:agn_removal}
We first removed AGN from our final sample based on the AGN sources identified in \cite{Roberts-Borsani2024} and \cite{Brooks2024}. Their selection criteria for AGN is based on broad+narrow component models to reproduce the H$\alpha$ profiles while the [OIII]$\lambda$5007 profile is reproduced by a single narrow component. Following this criterion, we removed {12} sources from the CEERS survey, 14 sources from the JADES survey, 1 source from the GLASS survey, and 2 sources from the GO-3073 program. Therefore in the following sections, we consider a final sample of {761} galaxies. We note that our sample might still have contamination from narrow obscured AGN but there is not a clear method to select them at high-$z$. We discuss this possible contamination in Appendix \ref{appendix:images}.

We selected a subsample based on the detection limits of the surveys. We called this subsample the complete sample. We assumed the 5$\sigma$ depth of 29.3 mag in the F200W for a point source in the CEERS survey \citep{Finkelstein2025} {which is the shallower survey we considered}. The complete sample includes {570} galaxies above those depths.
\begin{figure}[t!]
    \centering
    \includegraphics[width=\columnwidth]{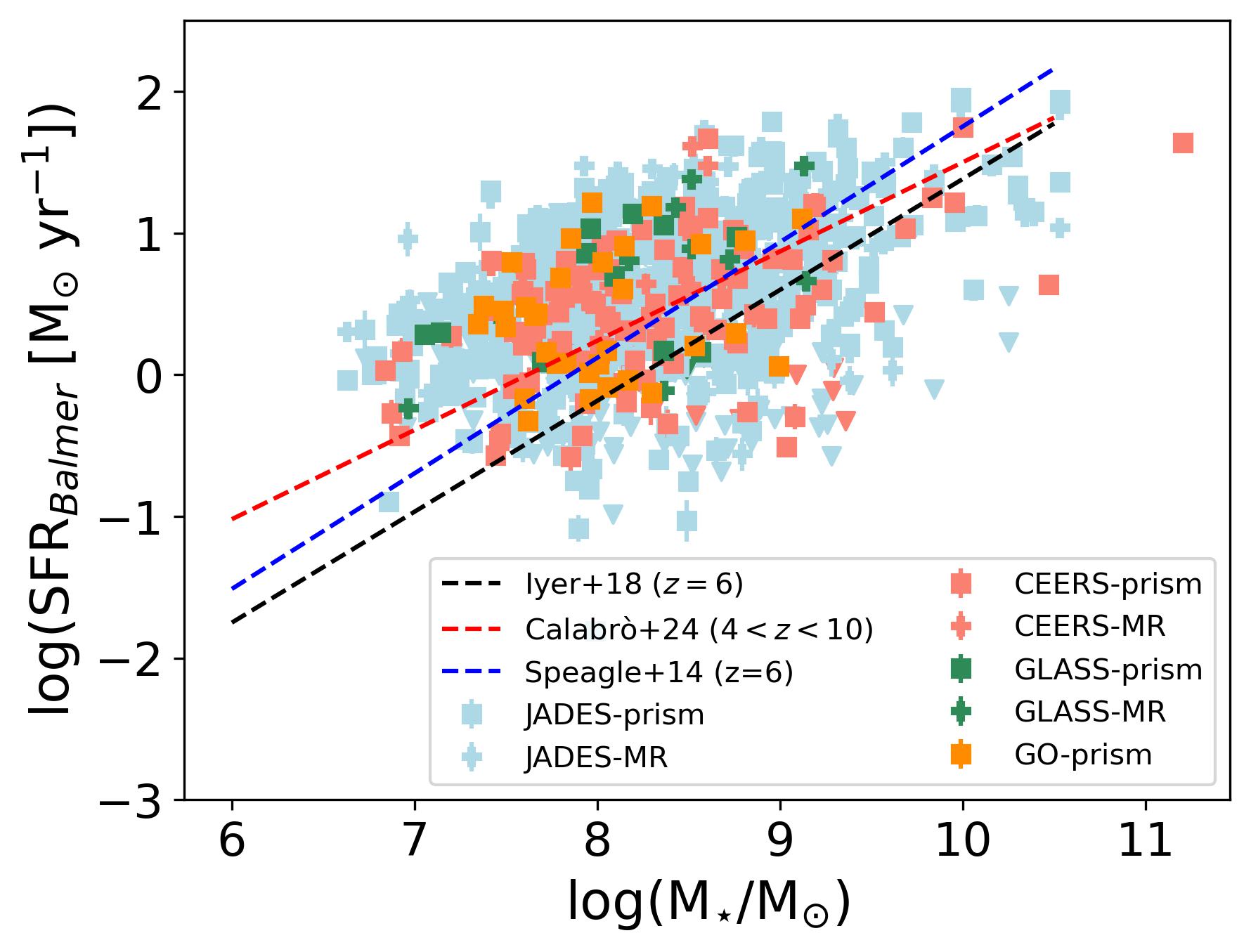}
    \caption{Distribution of the full sample along the stellar mass-SFR plane. {The dashed black and blue lines are the main-sequence of star-forming galaxies at $z=6$ from \cite{Iyer2018} and \cite{Speagle_2014}, respectively.} {While in dashed red line is represented the main-sequence at $4<z<10$ from \cite{Calabro2024sigmasfr}.} {The triangle symbols are upper limits based on \halpha or on \hbeta if an upper limit on \halpha is not estimated.}}
    \label{fig:MS}
\end{figure}

\begin{figure}[t!]
    \centering
    \includegraphics[width=\columnwidth]{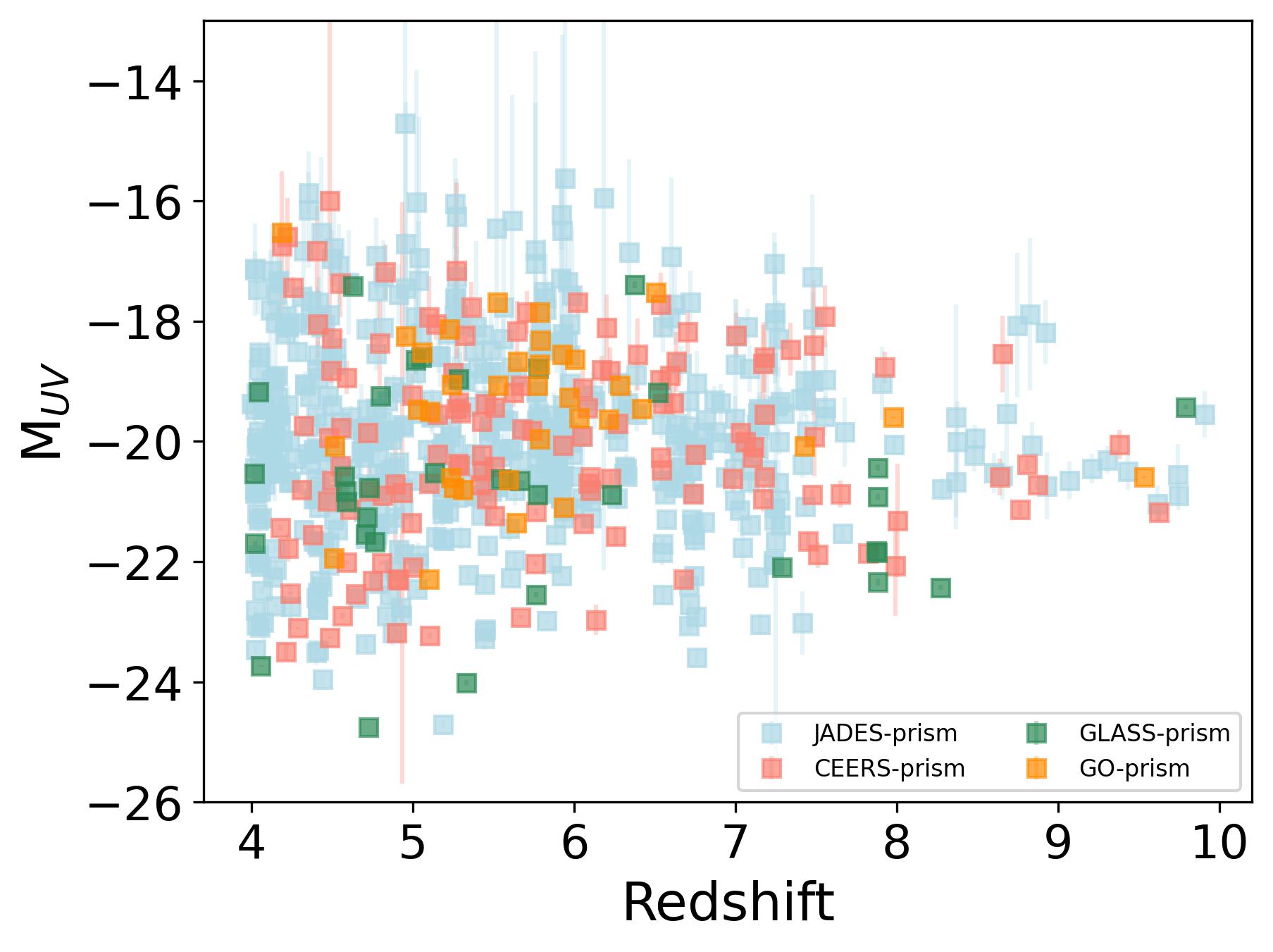}
    \caption{{Distribution of UV absolute magnitude as a function of redshift for the galaxies in the different surveys considered in our sample.}}
    \label{fig:z_Muv}
\end{figure}

\begin{figure*}[t!]
    \centering
    \includegraphics[width=\textwidth]{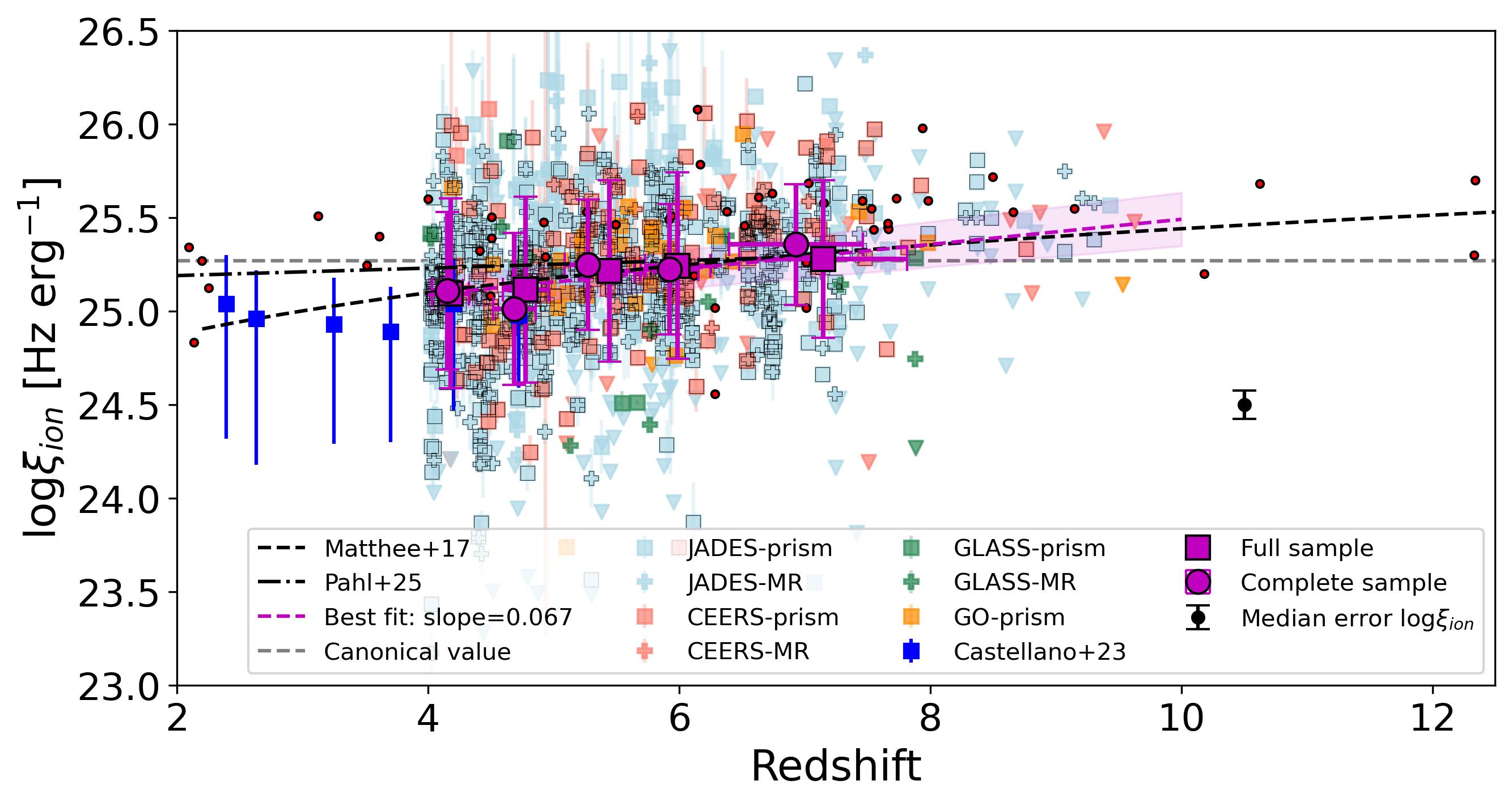}
    \caption{Redshift evolution of \xiion for the entire sample. The individual galaxies are in red, light-blue, green, and orange symbols, depending on the parent survey. The symbols with black edges are galaxies in the complete sample. The plus symbols are galaxies based on grating (MR) configuration while the square symbols are galaxies observed with prism configuration. The triangle symbols are upper limits. The magenta symbols are the median values of \xiion in equally populated bins of redshift. The magenta squares are the median values considering the full sample, while the magenta circles are the median values considering only the complete sample. 
    The blue symbols are average values from \cite{Castellano2023}.
    The red circles are individual galaxies or stacks from literature \citep{Stark2015civ,Nakajima2016,Marmol-Queralto2016,Bouwens2016,Matthee2017,Stark2017,Shivaei2018,Harikane2018,Vanzella2018,Faisst2019,Lam2019,DeBarros2019,Emami2020,Nanayakkara2020,Castellano2022,Marques-Chaves2022,Stefanon2022,Prieto-Lyon2023,Bunker2023jades,Rinaldi2024,Roberts-Borsani2024,Mascia2024ceers,Saxena2024,Lin2024,Alvarez-Marquez2024,Hsiao2024,Calabro2024,Vanzella2024,Zavala2024}. The magenta dashed line is the best fit of the median values of the full sample. The dashed line is the relation from \cite{Matthee2017} and the dotted-dashed line is the relation from \cite{Pahl2024}.
    The canonical value of $\log \xi_{\text{ion}}[\text{Hz} \ \text{erg}^{-1}] \approx 25.27 \ $, often assumed in reionization models, is highlighted in horizontal dashed grey line for reference. {The black symbol represents the median error of 0.07 dex in log\xiion. }}
    \label{fig:xi_z}
\end{figure*}
\subsection{Emission line measurements}\label{sec:lime}
To measure  \xiion in individual galaxies we need to estimate the luminosity of Balmer lines. In this case, we considered H$\alpha$ and H$\beta$ lines. For part of the analysis in the following sections, we also considered [OIII]$\lambda$5007 and the unresolved doublet [OII]$\lambda$3729 in the set of lines to measure.

For the CEERS survey, we considered the measurements included in the data Release {0.9} (Arrabal Haro et al. \textit{in prep.}). The set of measurements was performed using LiMe\footnote{\url{https://lime-stable.readthedocs.io/en/latest/}} \citep{Fernandez2024Lime} which is a library that provides a set of tools to fit lines in astronomical spectra.

For consistency, we used the same code to measure the line for the sample in the other surveys. We considered one gaussian to model the line profiles. For the galaxies in the GLASS survey and the GO-3073 program we corrected the fluxes for magnification using the lens model presented in \cite{Bergamini2023}. We also checked the flux calibration of the spectra based on the available photometry and we corrected the observed fluxes by a factor of $\sim$30\% of flux losses based on the median value in all the photometric bands. {This correction does not depend sensibly on the wavelength in agreement with what was found by \cite{Roberts-Borsani2024}}.

We corrected the line fluxes for dust reddening using the \cite{Calzetti2000} attenuation law. 
We considered the E(B-V) value from the SED fitting for galaxies, which is obtained as described in the following section {\ref{sec:sed-bagpipes}}. As a caveat, we did not consider the Balmer decrement in this project in order to have a homogeneous dust correction for all galaxies in the sample. {When comparing E(B-V) values from Balmer lines and SED fitting, they are different by a median (mean) factor of 0.47 (1.53). To avoid artificial trends induced by using different methods to estimate E(B-V) for the whole sample, we opted to consider the values from the SED fitting for the dust attenuation.} As we will show in Sec. \ref{sec:xi_measurment} and Appendix \ref{appendix:ha_hb}, we find a good agreement between the luminosities of \halpha and \hbeta which corroborates the robustness of our dust correction.

\subsection{SED fitting}\label{sec:sed-bagpipes}

We used BAGPIPES \citep{Carnall2018} {version 1.2.0} to estimate the physical parameters with the \cite{Bruzual_2003} stellar population models. We fixed the redshift to the spectroscopic redshift. We considered a delayed exponential $\tau$-model for the star formation history (SFH), where $\tau$ is the timescale of the decrease of the SFH. In the model, we consider an  age ranging from 1Myr to the age of the Universe at the observed redshift. We allowed the $\tau$ parameter to vary between 0.1 to 10 Gyr and the metallicity to vary up to 0.5Z$_\odot$ freely. The upper limit in stellar metallicity is based on the stellar mass-metallicity relation observed at $z=3.5$ for a stellar mass of 10$^{11}$\Msun \citep{Llerena2022,Stanton2024}. For the dust component, we considered the \cite{Calzetti2000} attenuation curve and let the A$_V$ parameter vary between $0-2$ mag. We also included a nebular component in the model, and we let the ionization parameter freely vary between {-4 and 0.} We used the same recipe for all the galaxies in all surveys using the available photometry. The observed photometry of galaxies in the GLASS survey and in the GO-3073 program is corrected using the same lensing model available in \cite{Bergamini2023}.

We also converted the dust-corrected $L(H\alpha)$ to SFR assuming the calibration from \cite{Reddy2022} as
$\text{SFR(Balmer)}=L(H\alpha)\times 10^{-41.67}$. This is the most suited calibration according to the typical subsolar metallicities expected for our galaxies at $z > 4$, and reflects the
greater efficiency of ionizing photon production in metal-poor stellar populations. The distribution of the sample along the main sequence at $z=6$ \citep{Iyer2018,Speagle_2014} is displayed in Fig. \ref{fig:MS}. {As a reference, we also show the main sequence from \cite{Calabro2024sigmasfr} that includes a wide range of redshifts between $z=4$ and $z=10$}. Our sample is scattered around the main sequence, and the galaxies cover $\sim 3$ dex in stellar mass and SFR. We note that there is no bias in stellar mass or SFR between the various subsamples depending on the considered survey. {Nonetheless, less massive sources with M$_{*}\lesssim10^{7.5}$\Msun\, tend to be above the main sequence, independently of the relation we use as reference.}

\subsection{UV luminosity density}

We determined the UV luminosity density by estimating the value of the rest-frame continuum at 1500\r{A} based on the SED model. For this purpose, we {considered a synthetic top-hat filter of bandwidth 100\r{A} centered at 1500\r{A} to convolve with the SED model.} 
Additionally, {we followed \cite{Pahl2024}} and estimated the uncertainty of the UV luminosity density by considering the mean uncertainty of the flux density in the observed filters covering the rest-frame from 1200 to 3000\r{A}. {In this way the error bars are motivated by nearby photometric constraints.} We did not estimate the rest-frame continuum at 1500\r{A} from the spectra because of the lack of good S/N in the continuum in grating spectra. To obtain reliable and homogeneous estimates, we used the available photometry for all sources. 

We corrected the obtained UV luminosity densities for dust reddening following the same procedure as for the emission line fluxes as described in Sec. \ref{sec:lime}. We finally converted the dust-corrected UV luminosity densities to UV absolute magnitudes which are analyzed in the following sections. {In Fig. \ref{fig:z_Muv}, we show the distribution of M$_{UV}$ vs redshift. Our sample covers a wide range of M$_{UV}$ ($\sim$ 8 mag) with a mean value of M$_{UV}=-20.00$ ($\sigma=1.61$). }{In Appendix \ref{appendix:balmer_flux} we show the distribution of the measured \halpha and \hbeta emission line flux as a function of both redshift and M$_{UV}$. We note that our complete sample includes galaxies with $z\lesssim7$ and M$_{UV}\lesssim-18$ which is the parameter space where our results are more robust.}

\subsection{Constraints on \xiion}\label{sec:xi_measurment}
We estimated the ionizing photon production efficiency in a standard way considering the model where \xiion is given by
\begin{equation}
    \xi_{ion}=\dfrac{N(H^0)}{L_{UV}},
\end{equation}
where $N(H^0)$ is the ionizing photon rate in s$^{-1}$ and L$_{UV}$ is the UV luminosity density at 1500\r{A}. In order to estimate the ionizing photon rate we use the dust corrected \halpha luminosity when available as:
\begin{equation}
    L(H\alpha)[erg s^{-1}]=1.36\times 10^{-12} N(H^0) [s^{-1}],
\end{equation}
which was derived from \cite{Leitherer1995} assuming no ionizing photons escape the galaxy (f$_{esc}$ = 0) and case B recombination. {For this conversion factor it is assumed a \cite{Salpeter1955} initial mass function (IMF) and a gas temperature of 10000K for the nebular emission in the evolutionary synthesis models.} Alternatively, we use the dust-corrected \hbeta luminosity as: 
\begin{equation}
    L(H\beta)[erg s^{-1}]=4.87\times 10^{-13} N(H^0) [s^{-1}].
\end{equation}
We checked for galaxies with \halpha and \hbeta available that both values are in very good agreement (see Fig. \ref{fig:xi_ha_hb}). We assume f$_{esc}$ = 0, meaning that all LyC photons are reprocessed into the Balmer lines. The derived \xiion value can be considered as a
lower limit since higher f$_{esc}$ will lead to a higher \xiion. If \halpha or \hbeta are detected with S/N$<3$, we put an upper 3$\sigma$ limit in the \xiion value for that galaxy. {We considered an upper limit only if $\sigma$ is lower than the detection limits of the surveys to avoid overestimating the upper limits. We also considered upper limits on \halpha or upper limits on \hbeta only when a limit on \halpha is not possible to estimate.} We obtained a mean value of log(\xiion [Hz erg$^{-1}$])=25.22 ($\sigma =0.42$dex) for the galaxies in the entire sample.  

\section{Results}\label{sec:results_all}
\subsection{Evolution of \xiion with redshift}\label{sec:redshift}
In Fig. \ref{fig:xi_z} we show the evolution of \xiion with redshift. The magenta squares are the median values in equally distributed redshift bins. For the median values, we considered the upper limits on \xiion as half their values {following the methodology in \cite{Calabro2024sigmasfr}. In data science, imputation of upper limits with half the detection limit is a common
approach, and simulation studies have found that this is a better choice compared to assuming the detection limit itself or a
zero value, as it introduces the least bias into
the estimates \citep{Beal2001}}. The errorbars are the observed 1$\sigma$ scatter within a given redshift bin. We note that there is a shallow trend where the \xiion increases with redshift from $4\leq z\leq 10$. Based on the median values, \xiion increases from log(\xiion [Hz erg$^{-1}$])={25.09} at {$z\sim 4.18$} to log(\xiion [Hz erg$^{-1}$])={25.28} at {$z\sim 7.14$}. The best linear fit of the median values considering the full sample leads to the relation {log(\xiion [Hz erg$^{-1}$])= ($0.06\pm0.012$)$\times z$+($24.82\pm0.07$).} For galaxies at {$z\gtrsim 7.14$}, the median value of \xiion is greater than the canonical value with a median value of log(\xiion [Hz erg$^{-1}$])={25.28}.

{Interestingly, our bins at $z\sim4-5$ are consistent within 1$\sigma$ with the values found in \cite{Castellano2023} with VANDELS galaxies at $2\leq z\leq 5$ using SED fitting to determine the \xiion values. They found no significant evolution of \xiion with redshift but combining both results they suggest an increase of \xiion from $z\sim2$ up to $z\sim10$.} This trend also agrees with the relation found in \cite{Matthee2017} based on \halpha emitters at $z\sim 2$ which seems to be valid up to $z\sim 12$, according to data from the literature which are included in Fig. \ref{fig:xi_z}. We also compare our results with the recent work from \cite{Pahl2024} with a sample of CEERS and JADES galaxies from $1.06 < z < 6.71$. We find that our results agree with the increase of \xiion with redshift but our median values are slightly lower than their observed trend, although consistent within the observed scatter. 

We also find that the median values are below the usually assumed canonical value of \xiion,  for the bins at { $z<7$ and then become higher than this value at higher redshift.} 
Given the large scatter, there are individual galaxies in each bin whose \xiion values are above the canonical values. We do not find a significant difference in the observed trend if we include only galaxies in the complete sample. We still observe the increase of \xiion with redshift up to {$z\sim 6.92$}. {We remark that the complete subsample extends up to $z\sim7.5$ which is the $z$ completeness of our study. As we show in Appendix \ref{appendix:balmer_flux}, the completeness in M$_{UV}$ is up to magnitude $\sim-18$ which is similar to the completeness of other spectroscopic samples \citep[e.g.,][]{Dottorini2024}.}

\subsection{Relation between \xiion and M$_{UV}$}\label{sec:xi_muv}
\begin{figure}[t!]
    \centering
    \includegraphics[width=\columnwidth]{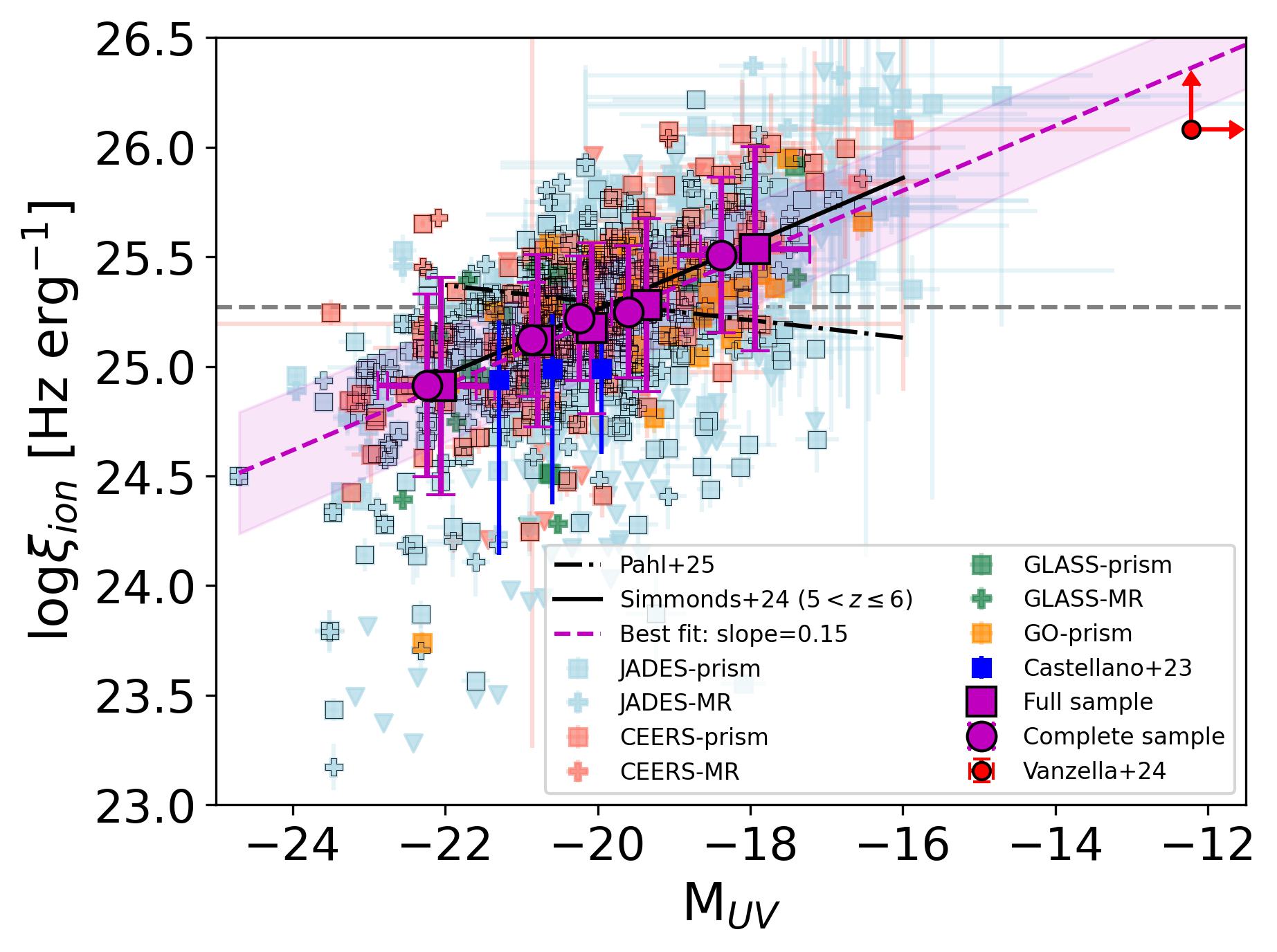}
    \caption{Relation between \xiion and UV absolute magnitude. Symbols are the same as in Fig. \ref{fig:xi_z}. The dotted dashed line is the relation found in \cite{Pahl2024} for galaxies at $1.06 < z < 6.71$. The black solid line is the relation found in \cite{Simmonds2024}. The red circle is the faint source from \cite{Vanzella2024}.}
    \label{fig:xi_MUV}
\end{figure}

In Fig. \ref{fig:xi_MUV} we show the relation between the \xiion and UV absolute magnitude. We find an increase of \xiion with UV magnitude which is also observed in individual galaxies. This is consistent with other works at $5\leq z\leq 7$ including galaxies in the Emission-line Galaxies and Intergalactic Gas in the Epoch of Reionization \citep[EIGER, ][]{Kashino2023} survey, CEERS, JADES \citep{Mascia2025} and also with works using photometry to estimate the \xiion values at comparable redshifts \citep[e.g][]{Simmonds2024} where we find a consistent slope. Our best fit of the median values is {log(\xiion [Hz erg$^{-1}$])= ($0.15\pm0.008$)$\times$M$_{UV}$+($28.17\pm0.17$)}. According to this trend, fainter galaxies tend to show higher \xiion values compared to brighter galaxies. In particular galaxies in the fainter bins ({median value $\gtrsim -19.36$ mag}) show $>$log(\xiion [Hz erg$^{-1}$])={25.28} higher than the canonical values. This trend is in the opposite direction compared with recent results from \cite{Pahl2024} {and \cite{Begley2025}} where they found that fainter galaxies are not the most efficient in producing ionizing photons but the brighter galaxies are. We also note that our relation perfectly agrees with the relation found in \cite{Simmonds2024} at $z\sim 5-6$. If we consider only the galaxies in the complete sample we find that the increasing trend is also found. Comparing with the literature, we find that our relation is in agreement with the \xiion value found recently in a faint source in \cite{Vanzella2024} with M$_{UV}\sim -12$, which suggests faint sources are indeed efficient producers of ionizing photons. 

\begin{table}[t!]
    \centering
    \caption{{Parameters of the best fits of the relation between \xiion and M$_{UV}$ at different redshift: log\xiion=a$\times$M$_{UV}$+b}}
    \begin{tabular}{c|c|c}
    Median $z$&a&b\\\hline
     4.42    &0.15$\pm$0.02 &28.15$\pm$0.48\\
     5.79    & 0.16$\pm$0.01&28.49$\pm$0.22\\
     7.45&0.21$\pm$0.02&29.76$\pm$0.36\\\hline\hline
    \end{tabular}
    \label{tab:xi_muv}
\end{table}

\begin{figure}[t!]
    \centering
    \includegraphics[width=\columnwidth]{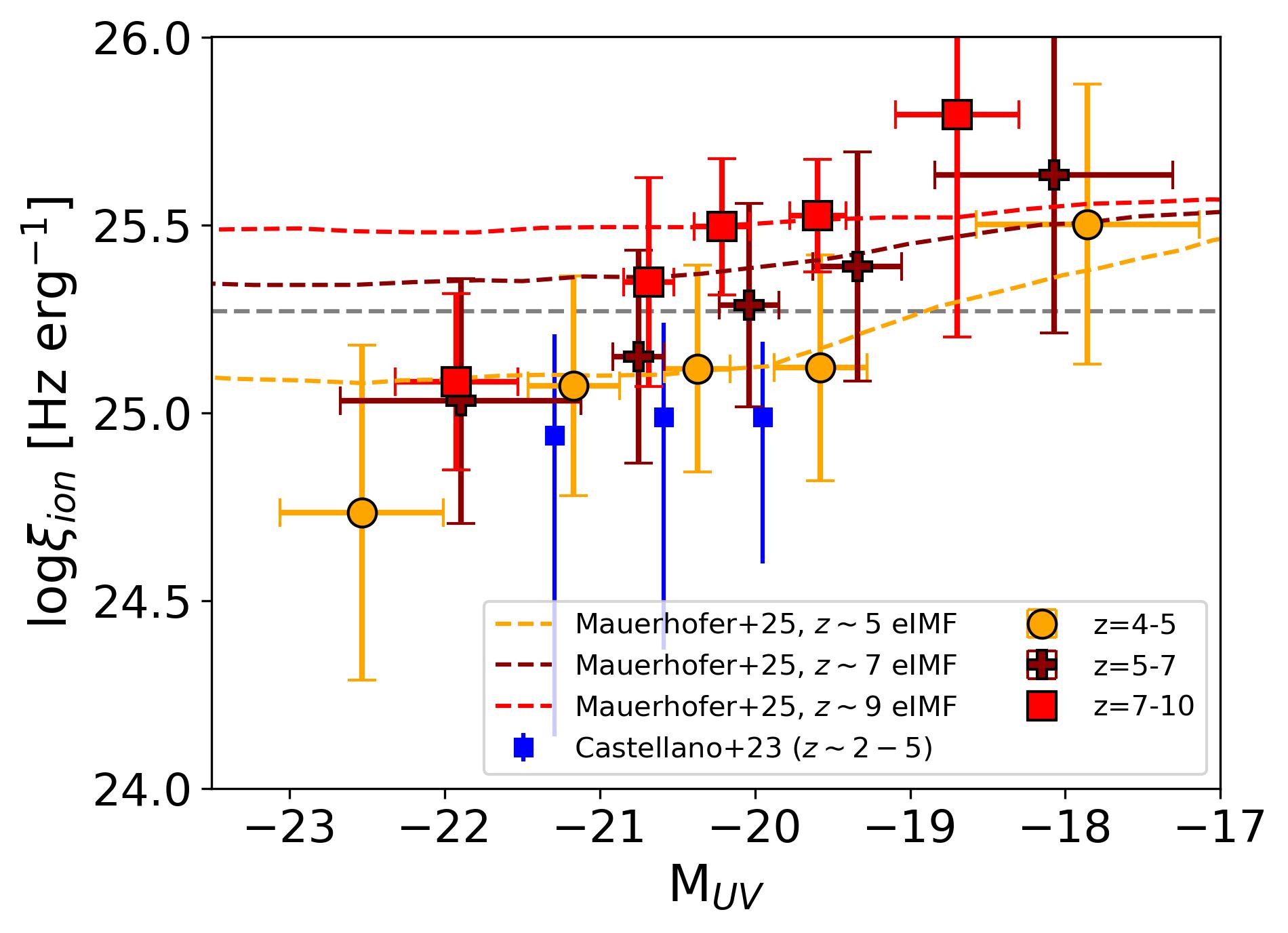}
    \caption{{Relation of \xiion with M$_{UV}$ in bins of redshift. The red, dark red, and yellow symbols are the median values in redshift bins. The dashed lines are models with an evolving IMF (eIMF) at different redshifts from \cite{Mauerhofer2025}. }{The dashed gray line is the canonical value for reference.}}
    \label{fig:xi_MUV_redshift}
\end{figure}

{In Fig. \ref{fig:xi_MUV_redshift} we explore if the relation of \xiion with M$_{UV}$ depends on the redshift. To this aim, we split the sample into three bins of redshifts: $z=4-5$ ({278 galaxies}), $z=5-7$ ({366 galaxies}), and $z=7-10$ ({117 galaxies}). We find that for a given M$_{UV}$, the galaxies with higher redshifts show higher \xiion values, which is in agreement with the redshift evolution found in Sec. \ref{sec:redshift}. {The slopes of the relations do not vary sensibly at different redshifts ({slopes between 0.15 and 0.21}). In Table \ref{tab:xi_muv} we present the parameters of the best linear fits depending on the median redshift.} 
We compare our results with the relations found in recent semi-analytic models with cold gas fractions and star-formation efficiencies obtained from hydrodynamical simulations \citep{Mauerhofer2025}. In particular, we compare our results with the models including an evolving IMF (eIMF) depending on redshift and metallicity, which becomes increasingly top-heavy with redshift or decreasing metallicity. Our results are consistent within 1$\sigma$ with the increase of \xiion with the increase of M$_{UV}$ found in the models at redshifts $z\sim5-7$  \citep[Fig. 14 in][]{Mauerhofer2025}. At $z\sim9$ the same models suggest a flattening of the relation at $\sim$log(\xiion [Hz erg$^{-1}$])={25.5} which is consistent within 1$\sigma$ with the values we obtained but not with the increasing trend we found in the bin with higher redshifts (median $z=7.45$). The lower number of galaxies in this bin and the wider redshift range could explain the differences with the models. In any case, our results are consistent with the models with a top-heavy IMF.}

\subsection{Total ionizing emissivity}\label{sec:total_emissivity}
\begin{figure}[t!]
    \centering
    \includegraphics[width=\columnwidth]{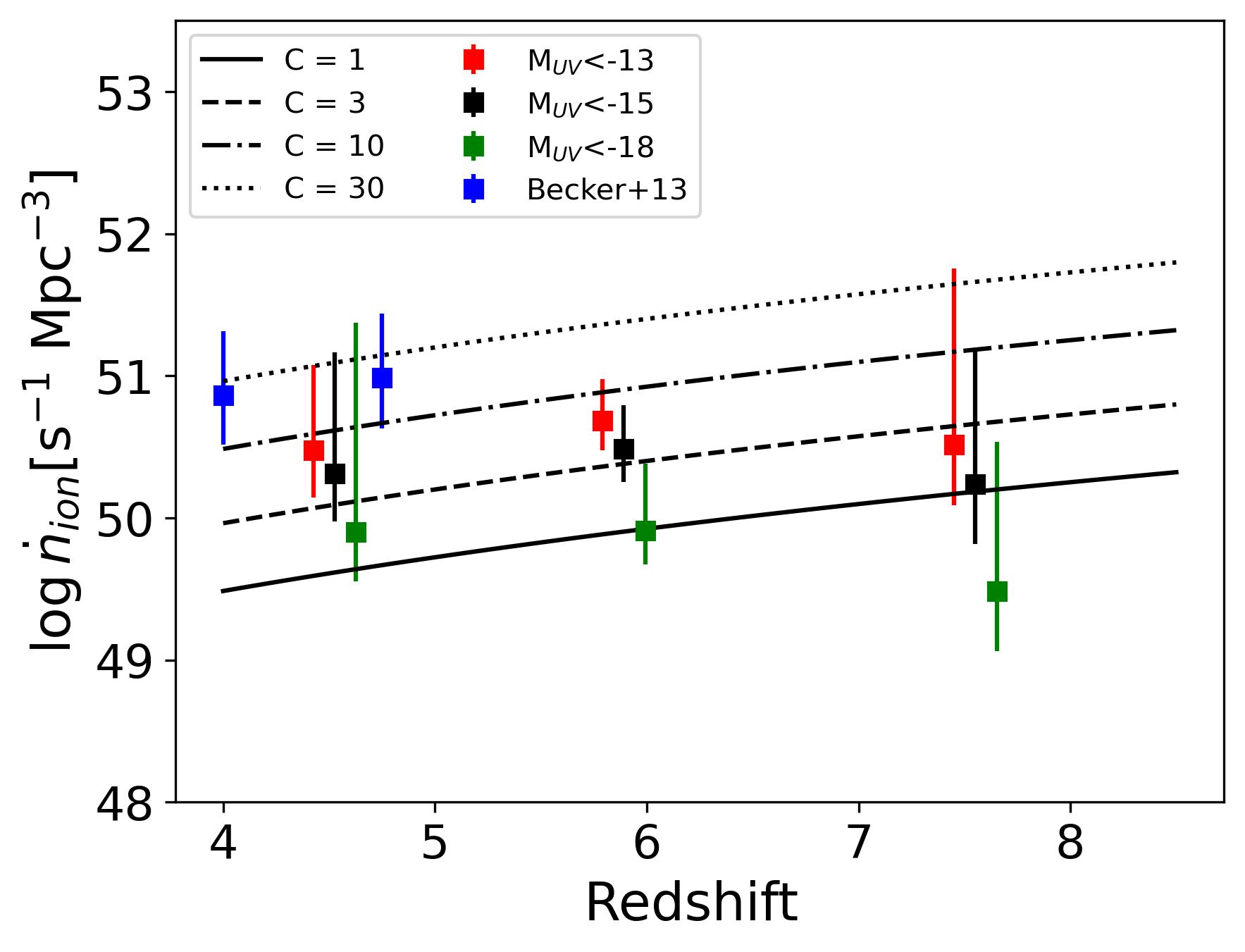}
    \caption{{Evolution of the total ionizing emissivity. The red, black, and green squares are our estimations after integrating from M$_{UV}=-23$ to M$_{UV}=-13$, M$_{UV}=-15$, M$_{UV}=-18$, respectively. The black and green squares are shifted by 0.1 and 0.2 in redshift, respectively, for better visualization. The different lines are IGM evolution models from \cite{Madau1999} for different ionized hydrogen clumping factors $C$. The blue squares are observational constraints obtained by observing the Ly$\alpha$ forest from \cite{Becker2013}.}}
    \label{fig:total_ionizing}
\end{figure}

{We use the relations reported in Table \ref{tab:xi_muv} to estimate the total ionizing emissivity $\dot{n}_{ion}$ at the median redshifts in each bin. To do this calculation, we considered the UV luminosity functions of \cite{Bouwens2021} and a median f$_{esc}=0.13$ as recently inferred  using indirect observational estimates at $z\sim5-9$ \citep{Mascia2023,Mascia2024ceers}. This value does not depend substantially on M$_{UV}$, at least in the magnitude range that can be currently explored.  We also considered a maximum value of log(\xiion [Hz erg$^{-1}$])={26} which is in agreement with models of young metal-poor stellar populations \citep{Raiter2010,Maseda2020} and simulations of PopIII populations \citep{Lecroq2025}. We integrated the luminosity function from M$_{UV}=-23$ to different faint limits and presents the result in Fig.
\ref{fig:total_ionizing}  with limits 
M$_{UV}=-13$ (plotted in red), M$_{UV}=-15$ (in black), and M$_{UV}=-18$ (in green). We performed 1000 Monte Carlo simulations assuming a normal distribution of the parameters of the UV luminosity functions and the best-fits in Table \ref{tab:xi_muv}, keeping the escape fraction as constant. The plotted values are the corresponding median values, and the errorbars are the 16th and 84th percentiles resulting from the simulations.}

{We match observational constraints obtained by observing the Ly$\alpha$ forest from \cite{Becker2013}. Our results are also consistent with IGM evolution models \citep{Madau1999} and do not support recent claims for a budget-crisis and a too early end of reionization \citep[e.g.][]{Munoz2024}. We also find that the contribution to the total ionizing emissivity of UV faint galaxies ($-15\leq$M$_{UV}\leq-13$) becomes more important at higher redshift, going from a contribution of $\sim30$\% at $z\sim4.4$ to a contribution of $\sim47$\% at $z\sim7.4$ \citep{Mascia2024ceers}. On the other hand, the contribution of UV bright galaxies becomes less important at higher redshift, going from a contribution of $\sim26$\% at $z\sim4.4$ to a contribution of $\sim9$\% at $z\sim7.4$. These results indicate that faint galaxies are the dominant sources of ionizing photons during the EoR.}


\subsection{Dependence of \xiion on the physical properties}\label{sec:properties}

\begin{figure}[t!]
    \centering
    \includegraphics[width=\columnwidth]{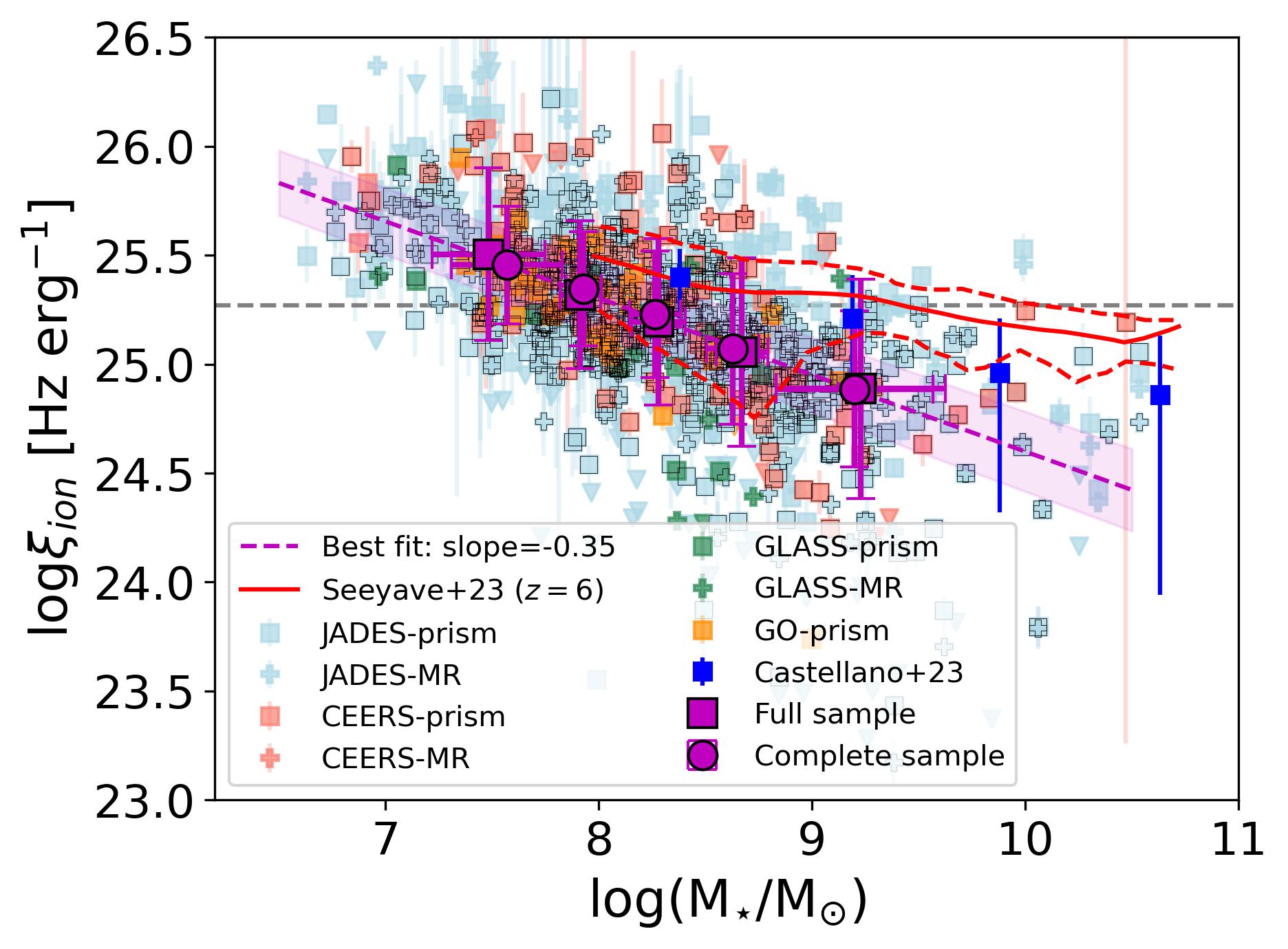}\\
    \includegraphics[width=\columnwidth]{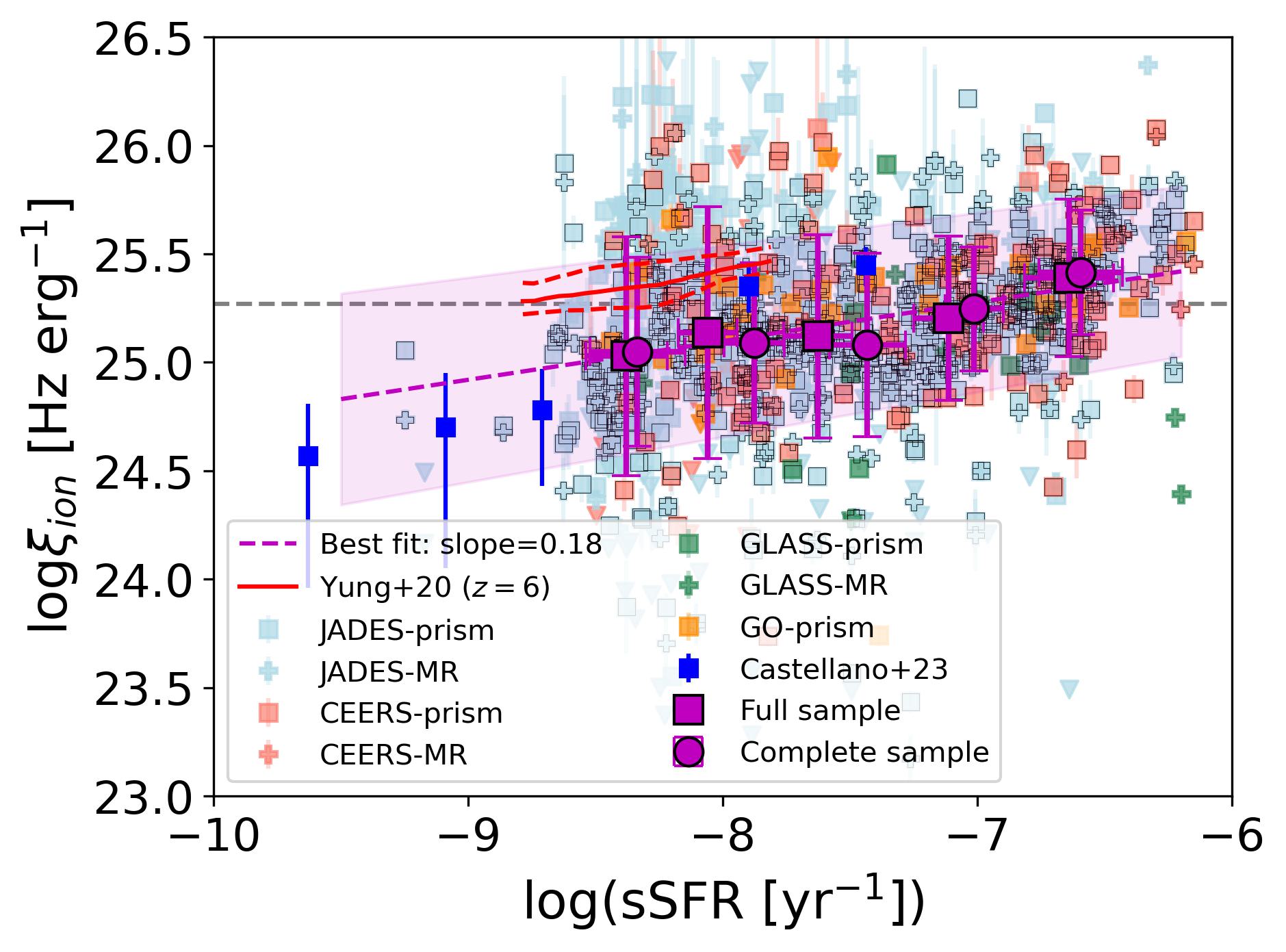}
    \caption{Relation of \xiion with physical properties of the galaxies. In the top panel, the relation with stellar mass, while in the bottom panel, the relation with the specific star-formation rate. Symbols are the same as in Fig. \ref{fig:xi_z}. In the top panel, the red solid line is the trend from the FLARES simulation at $z=6$ \citep{Seeyave2023} and the red dashed lines are their 3$\sigma$ scatter. In the bottom panel, the red solid line is the trend from simulations in \citep{Yung2020simulation} at $z=6$ and the dashed red lines are the 16th and 84th percentiles.}
    \label{fig:xi_mass}
\end{figure}

\begin{figure}[t!]
    \centering
    \includegraphics[width=\columnwidth]{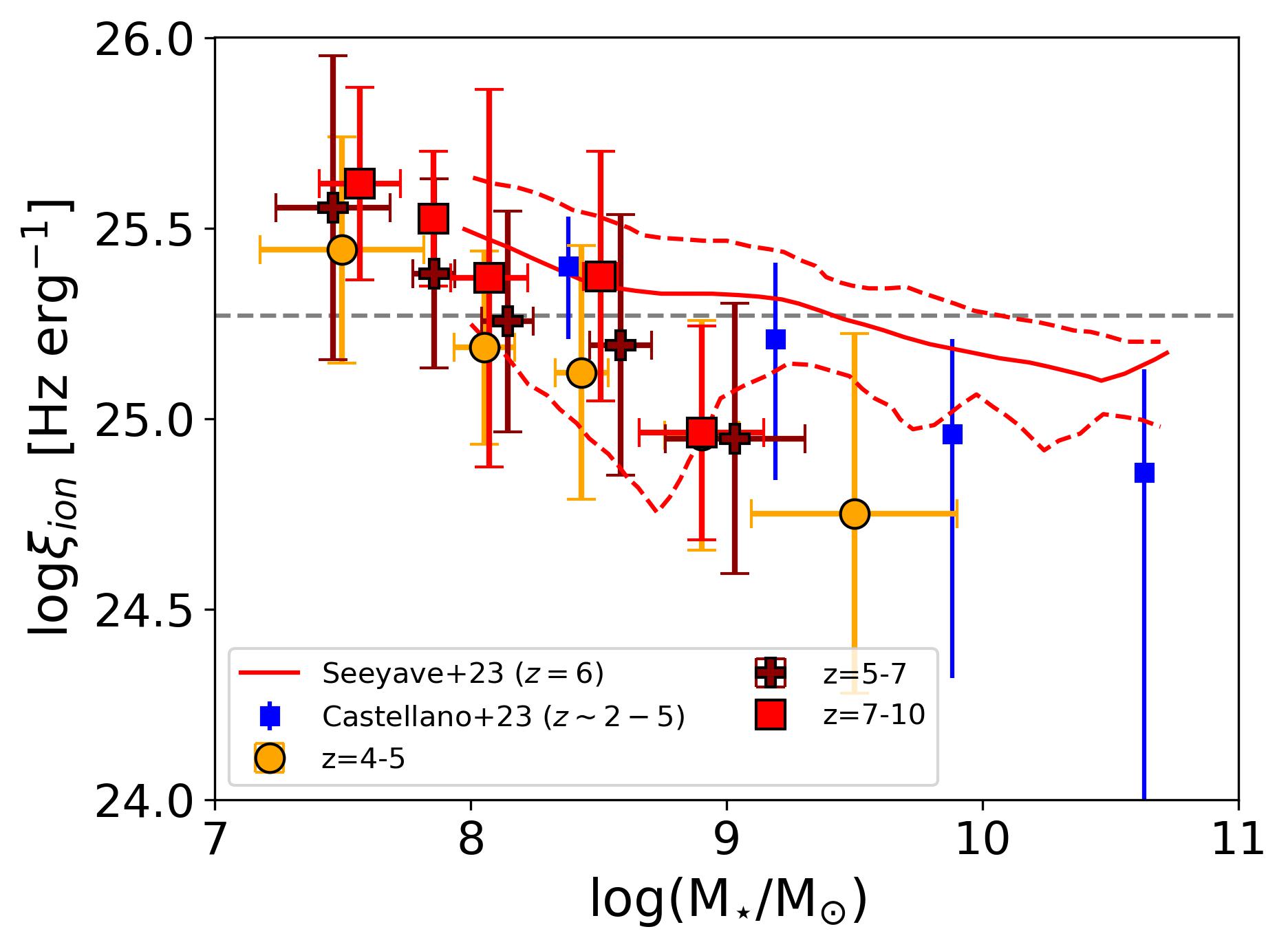}\\
    \includegraphics[width=\columnwidth]{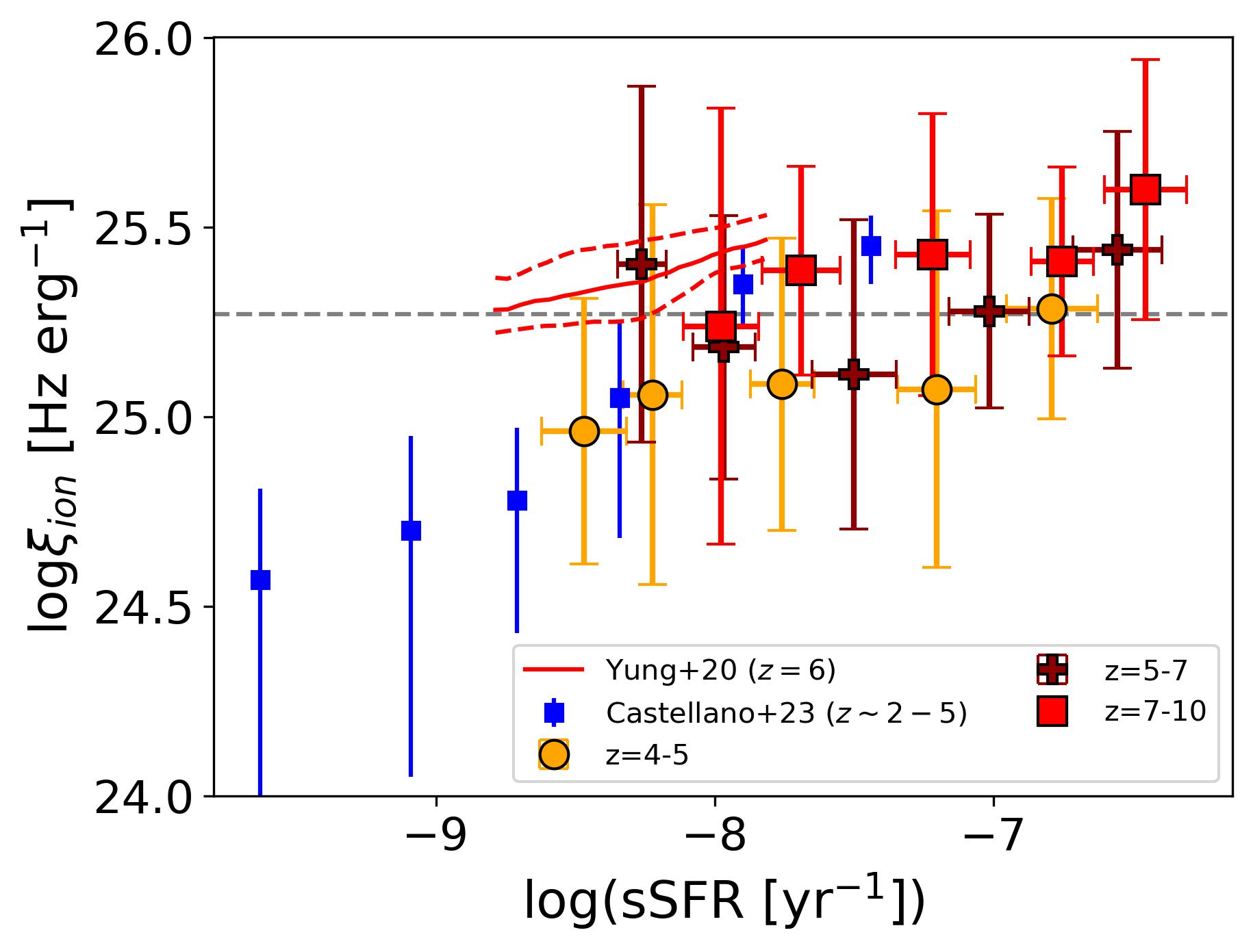}
    \caption{Relation of \xiion with physical properties of the galaxies in bins of redshift. The red, dark red, and yellow symbols are the median values in redshift bins.{The dashed gray line is the canonical value for reference.}}
    \label{fig:xi_mass_redshift}
\end{figure}

In this section, we present our results on the relation of \xiion with the physical properties obtained based on SED fitting. On the top panel in Fig. \ref{fig:xi_mass} we show the relation between \xiion with the stellar mass. Based on the median values in stellar mass bins, we find a clear decrease of \xiion with stellar mass. The best fit shown in Fig. \ref{fig:xi_mass} considers the median values. The best fit of the relation that we found is {log(\xiion [Hz erg$^{-1}$])= ($-0.35\pm0.01$)$\times$log(\Mstar/\Msun)+($28.12\pm0.11$).} 
This trend is in agreement with the one observed in VANDELS galaxies at lower redshifts \citep{Castellano2023}. We note that there is a small offset, {with the VANDELS sources showing slightly higher \xiion\ values for a given stellar mass:} however this could be an effect of the method used to estimate \xiion since we are using Balmer lines and in \cite{Castellano2023} they used SED modeling with BPASS (see more details in Sec. \ref{sec:EW}). Also, we highlight that in \cite{Castellano2023} the sample is complete for galaxies with stellar masses $>10^{9.5}$\Msun\, which could lead to the offsets observed on the lower mass bins.

This decreasing trend is also observed in \cite{Simmonds2024} using NIRCam photometry for a sample of $\sim 670$ galaxies at $z\sim 3.9-8.9$. Interestingly, the two bins with lower stellar mass ({median value $\lesssim$10$^{7.91}$\Msun}) show a median $>$log(\xiion [Hz erg$^{-1}$])={25.31} above the canonical value, which indicates that galaxies with lower stellar masses than this value tend to be efficient in producing ionizing photons. We also compare the trend we find with the results from simulations. In particular, we compare with the results from \cite{Seeyave2023} based on the FLARES simulation. They find a shallower decrease of \xiion with redshift, with higher values of \xiion for a given stellar mass. Our values are consistent with the 3$\sigma$ scatter of their relation, in particular for low stellar masses. The differences could be attributed to using BPASS models in the simulations. {According to \cite{Seeyave2023}, the decreasing trend of \xiion with stellar mass is due to the combined effects of increasing age and metallicity with increasing stellar mass, with metallicity likely playing a
bigger role (compared to age) due to the weaker evolution of age with stellar mass.}

On the bottom panel in Fig. \ref{fig:xi_mass} we show the relation between \xiion with the sSFR. Based on the median values in sSFR bins, we find a clear increase of \xiion with sSFR. The best fit of the relation that we found is {log(\xiion [Hz erg$^{-1}$])= ($0.18\pm0.04$)$\times$log(sSFR[yr$^{-1}$])+($26.53\pm0.30$).} This trend is also clear in individual galaxies. This trend is also in agreement with the one observed in VANDELS galaxies at lower redshifts \citep{Castellano2023} in the common sSFR range covered by the two studies. 
Similar to the relation with stellar mass, we note that the bin with higher sSFR (median value {$\sim$10$^{-6.64}$} yr$^{-1}$) shows a median log(\xiion [Hz erg$^{-1}$])={25.38} above the canonical value, which indicates that galaxies with higher sSFR than this value tend to be efficient in producing ionizing photons. We also note that considering the complete sample, galaxies with {$\sim$10$^{-6.59}$} yr$^{-1}$) tend to also show \xiion above the canonical values with a median value of log(\xiion [Hz erg$^{-1}$])={25.39}. We also compare our results with simulations from \cite{Yung2020simulation} and found a similar trend of increasing \xiion with increasing sSFR but we found an offset. Besides that fact, we found a similar slope to the one found in simulations.

In Fig. \ref{fig:xi_mass_redshift} we explore if the relations of \xiion with stellar mass and sSFR depend on redshift. To this aim, we split the sample into the same three bins of redshifts as in the previous section \ref{sec:xi_muv}. In Fig. \ref{fig:xi_mass_redshift} we show the median values of each bin. We find that for a given stellar mass, the galaxies with higher redshifts show higher \xiion values which is in agreement with the redshift evolution found in Sec. \ref{sec:redshift}. Similar results are found with sSFR with the galaxies in the bin $z=7-10$ showing the highest \xiion values. {The slopes of the relations do not vary sensibly at different redshifts ({slopes between -0.33 and -0.44 for the relation with stellar mass, and slopes between 0.05 and 0.18 for the relation with sSFR}), implying that the physical conditions leading to the photon production in galaxies remain essentially the same across cosmic epochs{, i.e. for the same metallicity and ages the efficiency is the same.}}

\subsection{Relation between \xiion and EW ([OIII])}\label{sec:EW}
\begin{figure}[t!]
    \centering
    \includegraphics[width=\columnwidth]{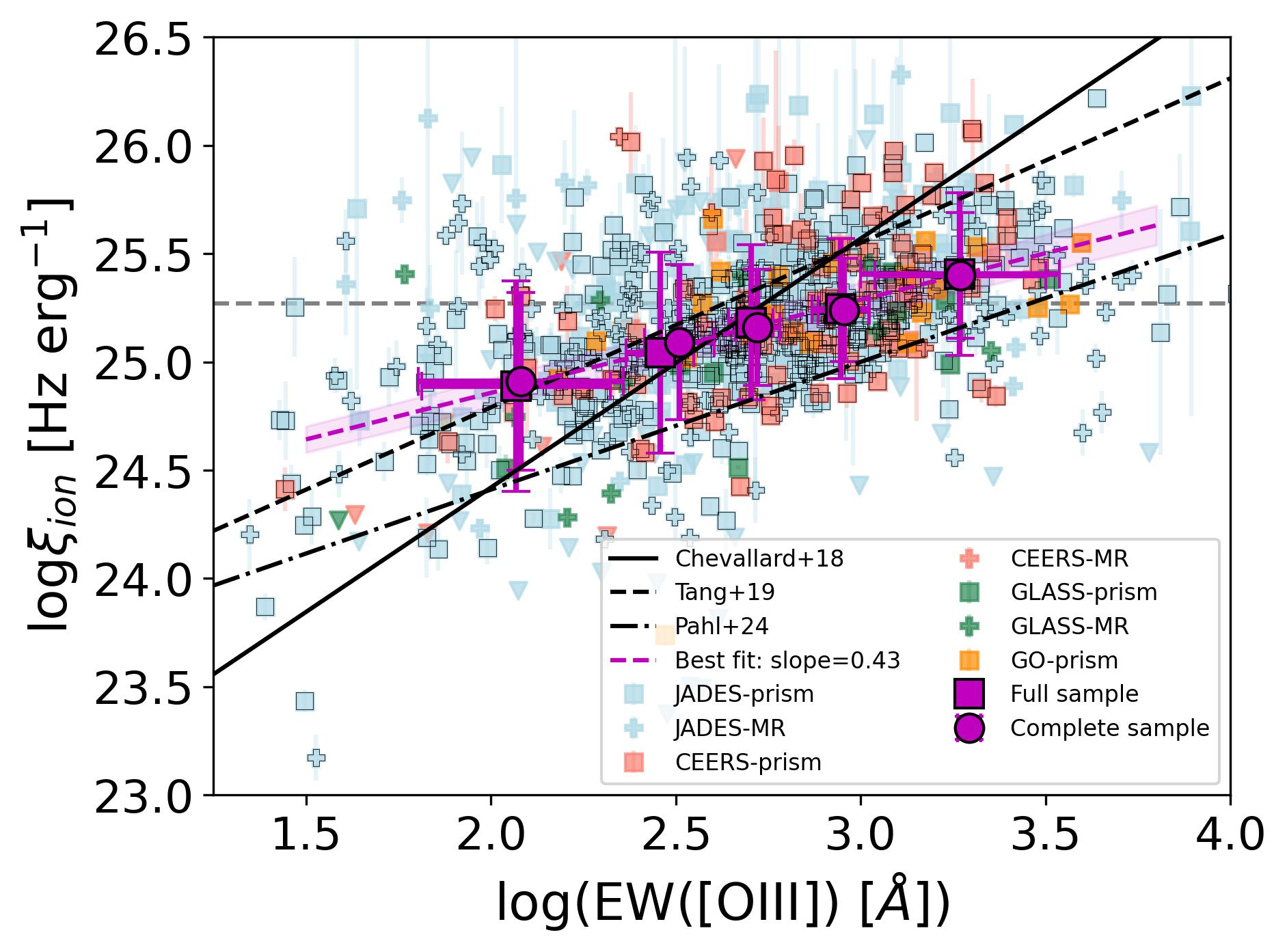}
    \caption{Relation between \xiion and EW([OIII]). Symbols are the same as in Fig. \ref{fig:xi_z}. The black solid line is the local relation from \cite{Chevallard2018}. The dashed line is the relation from \cite{Tang2019} at $1.3 < z < 2.4$. The dotted dashed line is the relation found in \cite{Pahl2024} for galaxies at $1.06 < z < 6.71$}
    \label{fig:xi_ewo3}
\end{figure}

{The EW([OIII]) has been often used as a proxy for \xiion. For example, \cite{Chevallard2018} found the relation marked with the solid line in Fig. \ref{fig:xi_ewo3} using a sample of 10 nearby analogs of primeval galaxies, which was then used in several works \citep{Castellano2023}. }
In Fig. \ref{fig:xi_ewo3} we show the relation between \xiion and the EW([OIII]) for our sample. For the estimation of the EWs we considered the continuum measured directly from the spectra based on the gaussian fitting. We find an increase of \xiion with the EW([OIII]) where the more efficient ionizing photon producers are the galaxies that show the higher EW([OIII]). In particular, we show that the galaxies in the bins with higher EWs (median value $\gtrsim$ {1800}\r{A}) show values of $>$log(\xiion [Hz erg$^{-1}$])={25.40} above the canonical values. The best fit of the median values results in the relation {log(\xiion [Hz erg$^{-1}$])= ($0.43\pm0.02$)$\times$log(EW([OIII])[\r{A}])+($23.99\pm0.05$).}
Compared to our trend, the \citep{Chevallard2018} relation is much steeper and tends to overpredict \xiion values for the most extreme cases of EW([OIII]) and at the same time underpredict the \xiion values in the cases with moderate EWs. A somewhat better agreement is found with the relation derived by  \cite{Tang2019} for a sample of $\sim$200 intense [OIII] emitters at $1.3 < z < 2.4$. We find a good agreement in the bins with lower EWs while in the more intense cases, our values tend to be lower than those predicted by these authors. Finally compared to the recent work by  \cite{Pahl2024}, we find slightly higher average values of \xiion for a given EW but the slope of the relation is consistent.  

We note that these offsets in the relations could also be the origin  for the discrepancies  observed in Fig. \ref{fig:xi_mass} when comparing our work to \cite{Castellano2023}. Their \xiion values are estimated using BEAGLE \citep{Chevallard2016} and are consistent with the \cite{Chevallard2018} relation which we found tends to overpredict the \xiion for a given EW, in particular in galaxies with high EWs. 

{Recently \cite{Laseter2024}  found that modest [OIII] emitters (EW$\sim$300-600\r{A}) may also show high values of \xiion as we also show in individual galaxies in Fig. \ref{fig:xi_ewo3}. One of the drivers of this effect could be the low metallicity in the systems where the EW([OIII]) is suppressed but the fluxes of Balmer lines remain elevated. We analyse the relation of \xiion with gas-phase metallicity in Sec. \ref{sec:metallicity}.  This dependency with gas-phase metallicity would limit the use of [OIII] as a sole tracer of high-$z$ efficient ionizing systems.}
\subsection{Relation of \xiion with O32 and gas-phase metallicity}\label{sec:metallicity}

\begin{figure}
    \centering
    \includegraphics[width=\columnwidth]{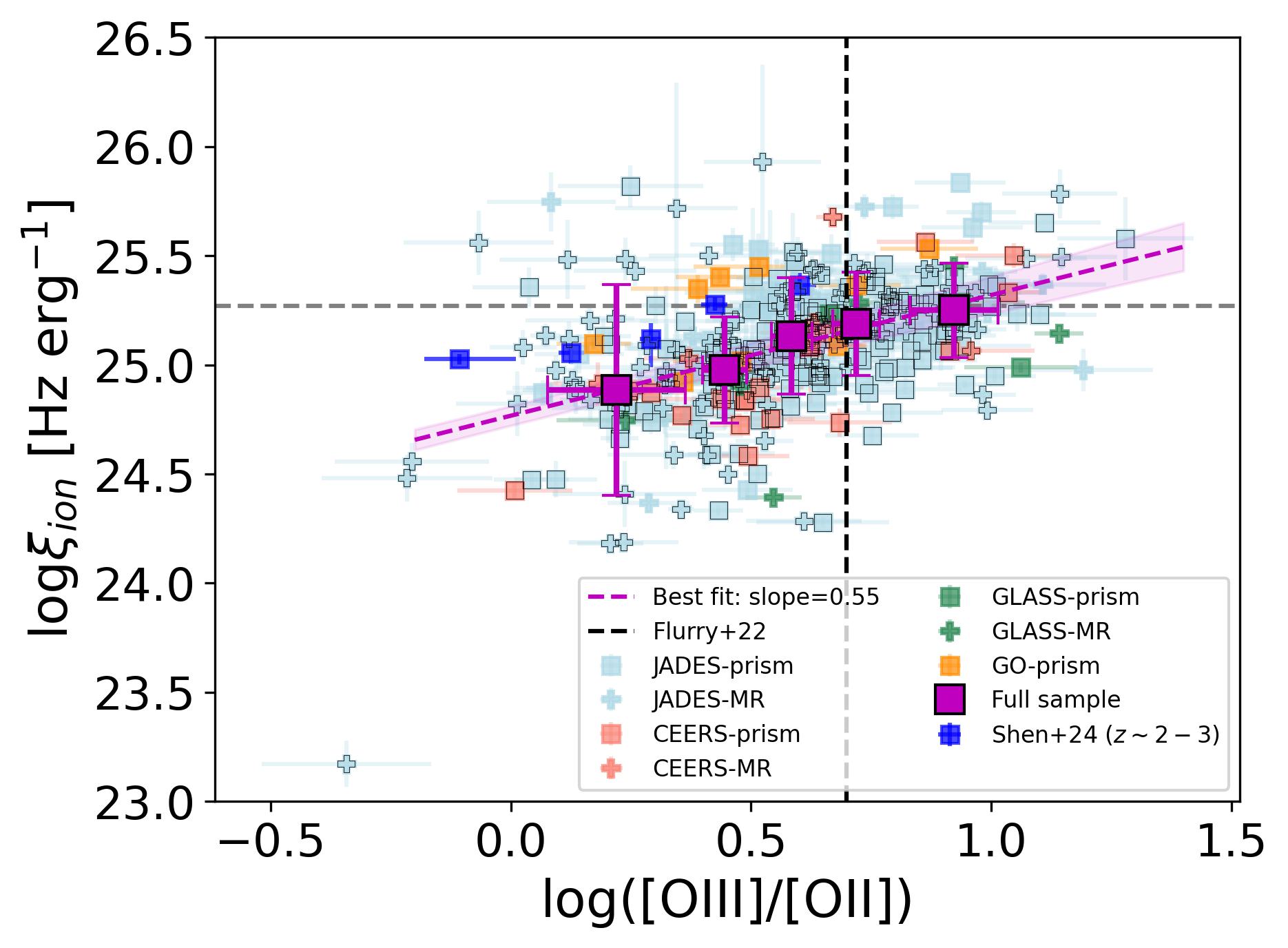}
    \caption{Relation between \xiion and O32. Symbols are the same as in Fig. \ref{fig:xi_z}. The blue squares are the median values from \cite{Shen2024} for galaxies at $z\sim 2-3$. {The vertical line is the demarcation line between strong and weak leakers from \cite{Flury2022}}.}
    \label{fig:xi_o32}
\end{figure}

In Fig. \ref{fig:xi_o32} we show the relation between \xiion and O32=log([OIII]/[OII]). We also include the limit O32$>$0.69 from \cite{Flury2022} to separate strong LyC leakers {found in the low redshift Lyman continuum survey galaxies}. We find an increasing trend where the galaxies with higher O32 values show the highest \xiion values. The best fit of the median values results in the relation {log(\xiion [Hz erg$^{-1}$])= ($0.55\pm0.07$)$\times$log([OIII]/[OII])+($24.77\pm0.04$).} 
The bin with the higher O32 values {($\sim$0.92)} is above the threshold for strong LyC leakers and it shows a median value log(\xiion [Hz erg$^{-1}$])=25.27, slightly above the canonical value. A similar trend was found in \cite{Shen2024} for galaxies at $z\sim 2-3$. We find a similar slope of the relation compared to their median values but our \xiion values are slightly lower for a given O32 value.  Since a high  O32 ratio is one of the proposed proxies for a high escape of Lyman continuum radiation \citep{Flury2022,Mascia2024ceers}, our findings could imply that galaxies with high ionizing photon production efficiency could also be those where the LyC escape is high which has implications in determining which sources contributed most to reionization. However, a high O32 is actually a necessary but not sufficient condition for a high escape fraction and some authors actually find that at least for some classes of galaxies (e.g., Ly$\alpha$ emitters), production and escape of ionizing photons are anticorrelated \citep{Saxena2024}. We plan to investigate these links further in future works.

\begin{figure}
    \centering
    \includegraphics[width=\columnwidth]{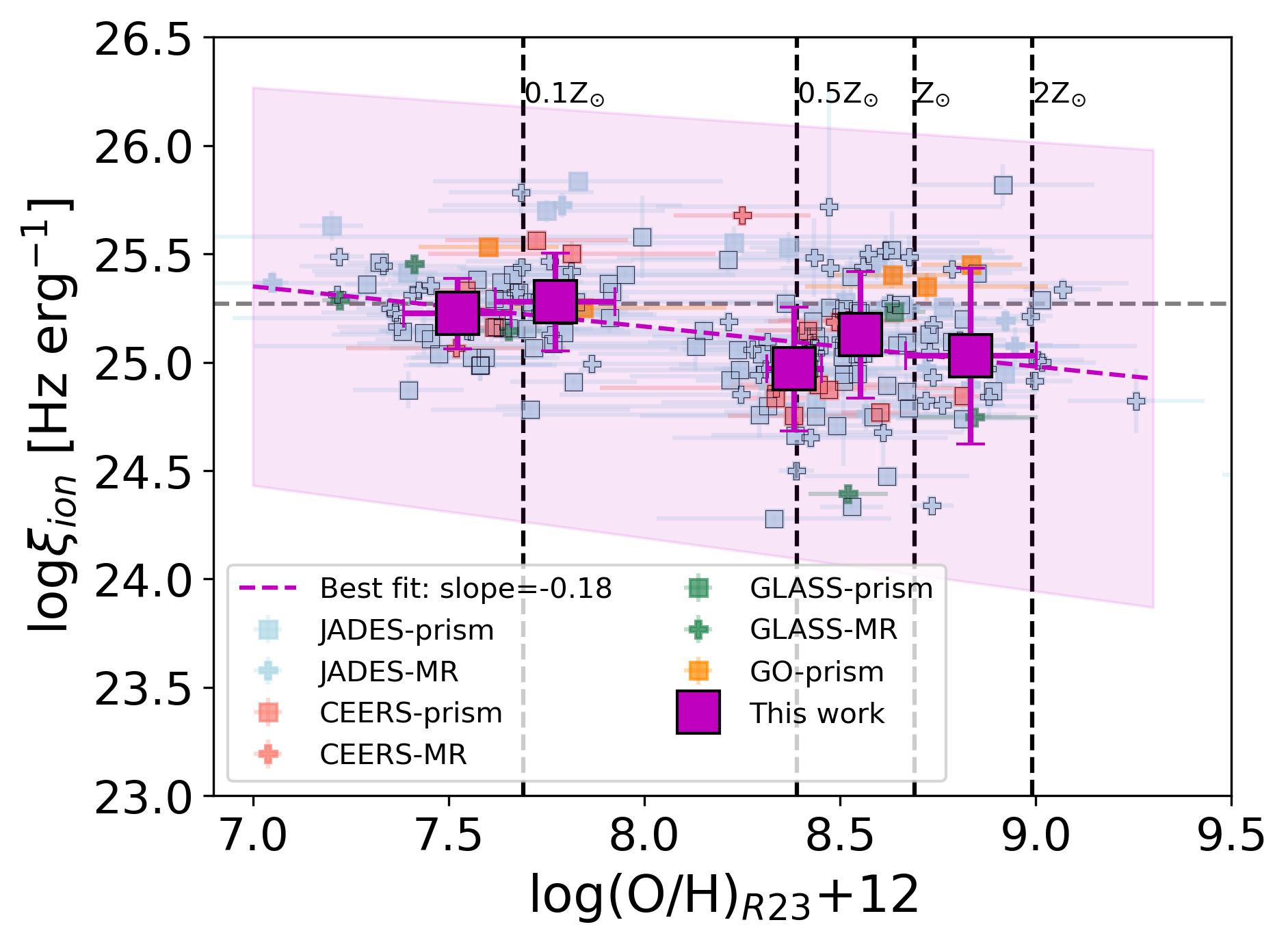}
    \caption{Relation between \xiion and gas-phase metallicity. Symbols are the same as in Fig. \ref{fig:xi_z}. }
    \label{fig:xi_oh}
\end{figure}

In Fig. \ref{fig:xi_oh} we show the relation between \xiion and gas-phase metallicity. We estimate the metallicity from the R23=($[O III]\lambda\lambda$4959,5007+$[O II]\lambda\lambda$3727,3729)/H$\beta$ calibration presented in \cite{Sanders2024} for galaxies at $z = 2.1-8.7$. Given that this relation is bivaluated we also used the O32 calibration in \cite{Sanders2024}. We consider as metallicity estimation the metallicity from the R23 calibration that is closer to the value from the O32 calibration. We find a shallow decreasing trend with metallicity, with the metal-poor galaxies showing the higher \xiion values. The best fit of the median values results in the relation {log(\xiion [Hz erg$^{-1}$])= ($-0.18\pm0.08$)$\times$(log(O/H)+12)+($26.64\pm0.69$).} 
According to the best fit of the median values, galaxies with metallicities $\lesssim$10\% solar tend to show \xiion values above the canonical value and are efficient in producing LyC photons. {We note that this relation is consistent with a flat slope at 3$\sigma$ and a larger sample would be needed to better constrain the dependency of \xiion on metallicity.}

\section{Summary and Conclusions}\label{sec:conclusions}
We selected a sample of {761} galaxies at $z = 4-10$ with NIRSpec spectra in four JWST surveys (CEERS, JADES, GLASS, and GO-3073) in this large study of \xiion conducted via spectroscopy. We estimated their physical properties using the available JWST and HST photometry for performing SED fitting using BAGPIPES \citep{Carnall2018} assuming a delayed exponential model for the SFH. We constrained the \xiion values based on Balmer lines (\halpha or \hbeta). We estimated the gas-phase metallicity using R23 and O32 calibrations for high-$z$ sources. We investigated the evolution of \xiion with redshift and the relations with the physical properties of the galaxies. Our main results are the following: 
\begin{itemize}
    \item We find an evolution of the ionizing photon production efficiency of star-forming galaxies with higher values of \xiion at higher redshifts. The evolution is mild, with the best linear relation given by {log(\xiion [Hz erg$^{-1}$])= ($0.06\pm0.012$)$\times$z+($24.82\pm0.07$).} This trend is consistent with other results at lower redshifts. According to this relation, a median value of log(\xiion [Hz erg$^{-1}$])={25.12} is inferred at $z=4.5$, while a median value of log(\xiion [Hz erg$^{-1}$])={25.32} is inferred at $z=7.5$. 

        \item {We find an increase of \xiion with increasing M$_{UV}$. The best-fit results in the relation {log(\xiion [Hz erg$^{-1}$])= ($0.15\pm0.008$)$\times$M$_{UV}$+($28.17\pm0.17$).} We also evaluated the  \xiion vs M$_{UV}$ relation in separate redshift bins and found that the slopes of the different relations stay constant, although with increasing offset. The relations are consistent with the predictions of recent models with an evolving IMF  becoming top-heavy at higher redshifts and lower metallicity.}
    \item {Assuming our values of  \xiion as a function of M$_{UV}$ and redshift, and a mean escape fraction of 0.13 as recently derived by \cite{Mascia2025}, we calculate the evolution of the 
    total ionizing emissivity. Our results show that galaxies can sustain reionization provided the clumpiness factor does not exceed 10 and are consistent with low redshift constraints by the Ly$\alpha$ forest.  We also find that UV faint galaxies ($-15\leq$M$_{UV}\leq-13$) are the main contributors of ionizing photons in the early stages of reionization.}
    
    \item We find a decrease of \xiion values with increasing \Mstar with the best-fit resulting in the relation {log(\xiion [Hz erg$^{-1}$])= ($-0.35\pm0.01$)$\times$log(\Mstar/\Msun)+($28.12\pm0.11$)}. 
    We find an increase of \xiion with increasing sSFR similar to other studies at lower redshifts. The best-fit results in the relation {log(\xiion [Hz erg$^{-1}$])= ($0.18\pm0.04$)$\times$log(sSFR[yr$^{-1}$])+($26.53\pm0.30$).} {This indicates that low-mass, faint UV, and with high levels of sSFRs galaxies tend to be efficient in producing ionizing photons.} The slopes of the above relations do not significantly change with redshift, implying that the conditions for photon production do not change and that the redshift evolution is only due to the different statistical properties of populations at each redshift.

    \item We find an  increase of \xiion with EW(O[III]). The best-fit results in the relation {log(\xiion [Hz erg$^{-1}$])= ($0.43\pm0.02$)$\times$log(EW([OIII])[\r{A}])+($23.99\pm0.05$)}. Compared to our trend, the widely used relation found by \cite{Chevallard2018} is much steeper and tends to overpredict \xiion values for the most extreme cases of EW([OIII]) and contrary, underpredict the \xiion values in the cases with moderate EWs. 
    We find instead a better agreement with the relation proposed by  \cite{Tang2019}, at least in the bins with lower EWs while in the more intense cases, our \xiion\ values are lower. 
    \item We find an increase of \xiion with O32 ratio. The best-fit results in the relation {log(\xiion [Hz erg$^{-1}$])= ($0.55\pm0.07$)$\times$log([OIII]/[OII])+($24.77\pm0.04$).} Since O32 is one of the most used proxies for a high escape fraction of Lyman continuum photons, this could imply that leakers could also be efficient in producing ionizing photons in contrast with some previous findings.
    \item Finally we find a decrease of \xiion with gas-phase metallicity. The best-fit results in the relation {log(\xiion [Hz erg$^{-1}$])= ($-0.18\pm0.08$)$\times$(log(O/H)+12)+($26.64\pm0.69$).} The rather shallow relation would indicate that metallicity does not play such a key role in determining the photon production efficiency.
\end{itemize}
Overall, we find that faint low-mass galaxies with high levels of sSFRs present the best conditions for an efficient production of ionizing photons, while the low metallicity seems to play a more marginal role in setting such conditions, given the shallow trend found. We also find indications that galaxies with high photon production efficiency might also be those where the conditions for high leakage of such photons are found since they share properties that are similar to the low redshift Lyman continuum leakers, namely high O32, faint UV magnitudes, and low-stellar masses.  Such galaxies could then be the main responsible for cosmic reionization: we plan to further investigate the link between ionizing photon production and escape in a follow-up work.  In general, although we do find that high redshift galaxies have higher \xiion compared to the low redshift sources with similar properties, our median values for the galaxy population during the EoR are not as extreme as those found by some other authors \citep[e.g.,][]{Maseda2020,Prieto-Lyon2023,Atek2024,Saxena2024}. In agreement with other authors \citep[e.g.,][]{Simmonds2024}{,
our} results therefore do not support some recent claims that we might have a budget-crises, i.e. that the total ionizing photons generated from galaxies are much higher than previously thought, which together with significant escape fractions could provide enough photons to end reionization too early, and in contrast to Lyman $\alpha$ forest results \citep{Munoz2024}. 

\begin{acknowledgements}    
    {We thank the anonymous referee for the detailed review
and useful suggestions that helped to improve this paper.} We wish to thank all our colleagues in the CEERS collaboration for their hard work and valuable contributions to this project. We thank Pietro Bergamini for providing us with the magnification factors for the lensed sources. MLl acknowledges support from the INAF Large Grant 2022 “Extragalactic Surveys with JWST” (PI L. Pentericci), the PRIN 2022 MUR project 2022CB3PJ3 - First Light And Galaxy aSsembly (FLAGS) funded by the European Union – Next Generation EU{, and INAF Mini-grant "Galaxies in the epoch of Reionization and their analogs at lower redshift" (PI M. Llerena).}
    This work is based on observations made with the NASA/ESA/CSA {\it James Webb Space Telescope (JWST)}. The JWST data presented in this article were obtained from the Mikulski Archive for Space Telescopes (MAST) at the Space Telescope Science Institute. The specific observations analyzed are associated with program JWST-GO-3073 and can be accessed via \href{https://doi.org/10.17909/4r6b-bx96}{DOI}.  We acknowledge support from INAF Mini-grant "Reionization and Fundamental Cosmology with High-Redshift Galaxies".
    This work has made extensive use of Python packages astropy \citep{astropy:2018}, numpy \citep{harris2020}, Matplotlib \citep{Hunter:2007} and LiMe \citep{Fernandez2024Lime}.
\end{acknowledgements}

%
%
\bibliographystyle{aa}
\bibliography{main}

\begin{appendix}
\section{Possible obscured AGN}\label{appendix:images}
\begin{figure}[t!]
    \centering
    \includegraphics[width=\columnwidth]{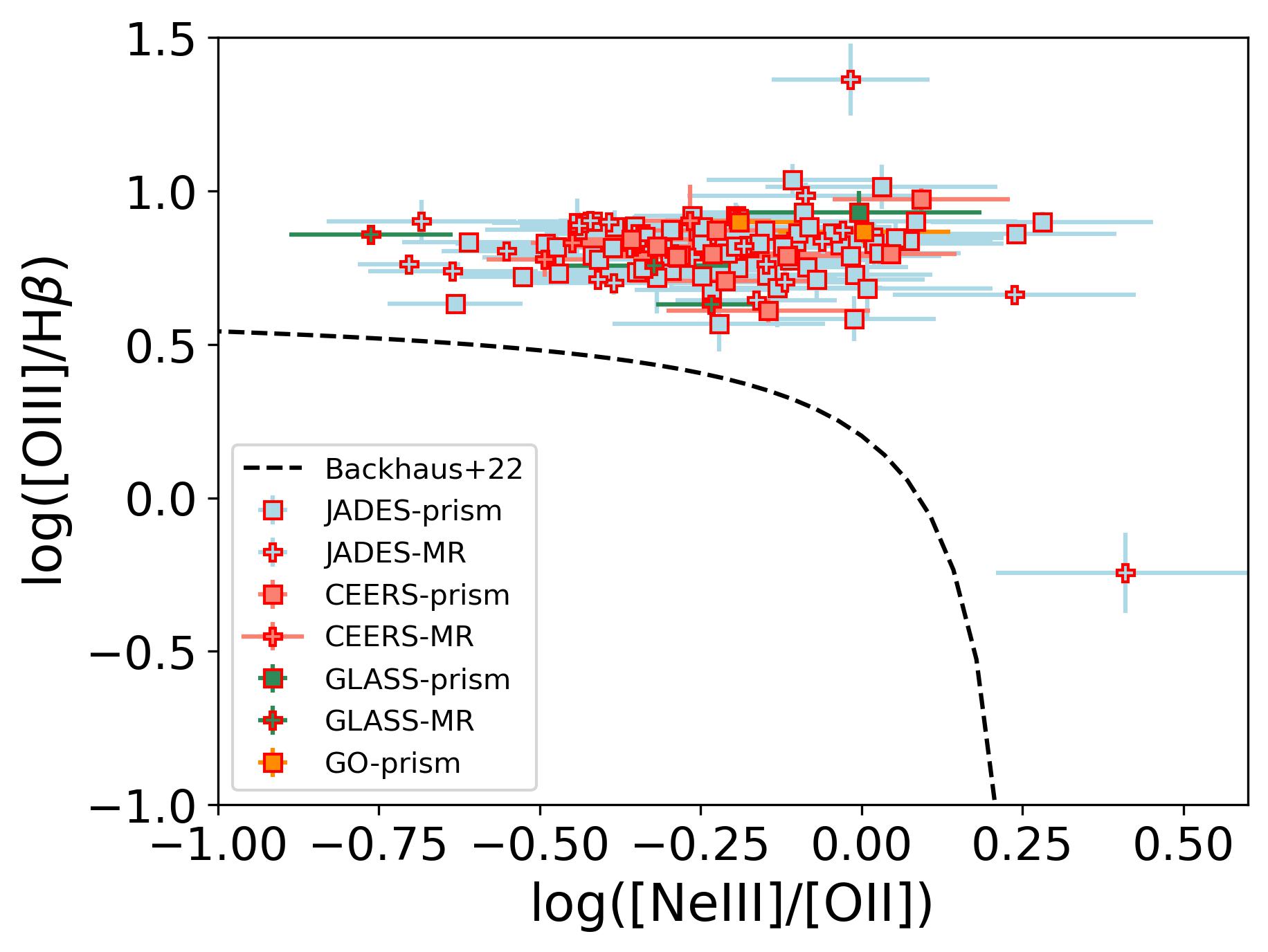}
    \caption{OHNO diagram for the galaxies in our sample. The symbols are the same as in Fig. \ref{fig:xi_z}. The dashed line is the demarcation line from \cite{Backhaus2022} to separate star-forming galaxies from AGN. AGN region is above this line based on this diagnostic. Based on this diagnostic, AGN candidates are shown as symbols with red edges.}
    \label{fig:OHNO}
\end{figure}

\begin{figure}[t!]
    \centering
    \includegraphics[width=\columnwidth]{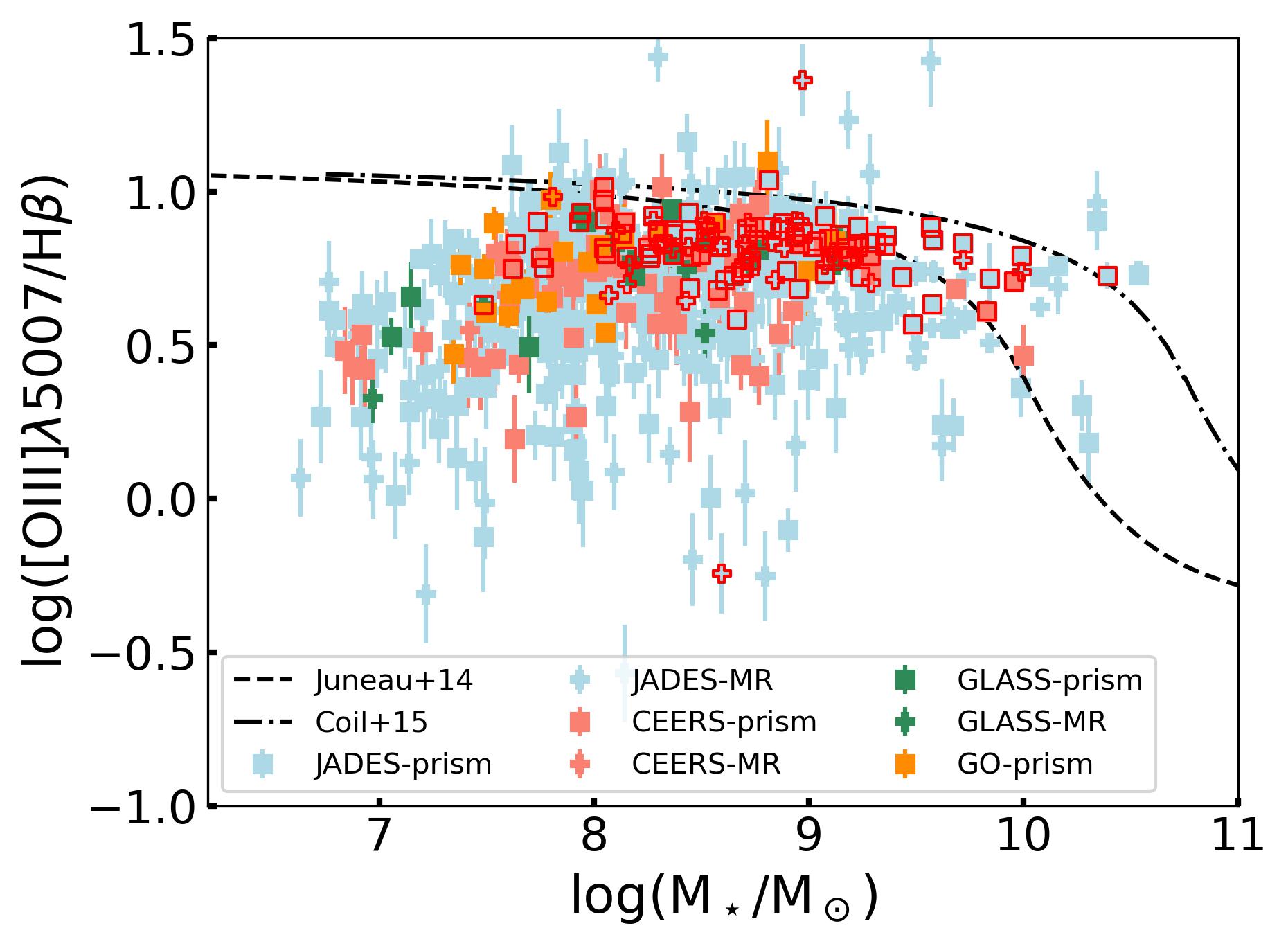}
    \caption{MEx diagram for the galaxies in our sample. The symbols are the same as in Fig. \ref{fig:xi_z}. AGN candidates from Fig. \ref{fig:OHNO} are shown as symbols with red edges. The black dashed line indicates the demarcation between star-forming galaxies and AGN, according to \cite{Juneau2014}. The black dashed-dotted line is the demarcation at $z\sim 2$ \citep{Coil2015}.}
    \label{fig:Mex}
\end{figure}

In Sec. \ref{sec:agn_removal} we removed unobscured AGN with broad Balmer lines from our sample galaxies at $z=4-10$. However, we note that our sample can still have a contribution of obscured narrow AGN. To analyze this possible contamination, in this section, we investigate the location of galaxies in some emission-line diagnostics. In particular, we analyze the OHNO diagram proposed in \cite{Backhaus2022} to separate galaxies ionized by massive stars from galaxies ionized by AGN. We analyze the subsample of galaxies with detected [NeIII]$\lambda$3870 and the blended [OII]$\lambda\lambda$3726,3728 line. In Fig. \ref{fig:OHNO} we show the location of this subsample in the OHNO diagram. We note that all these galaxies are above the demarcation line which indicates they could be  AGN. However, we remark that star-forming galaxies are also found in this locus, as can be seen in Fig. 6 in \cite{Backhaus2022}. Therefore, it is not clear that this subsample are AGN based only on this diagnostic diagram. For this reason, we analyze an alternative diagram that can be applied to this subsample. We choose the Mass-Excitation \citep[MEx, ][]{Juneau2014} diagram, which is shown in Fig. \ref{fig:Mex}. Based on this diagnostic diagram, we note that most of the galaxies in our sample are star-forming galaxies. The subsample of AGN candidates based on the OHNO diagram is also in the locus of star-forming galaxies in the MEx diagram. 
 Only a few candidates are above the demarcation line from \cite{Juneau2014} but not above the one from \cite{Coil2015} at $z\sim 2.3$. 
 We note however that the MeX diagram is not calibrated at the redshifts considered in this paper \citep{Coil2015}.

As a conclusion, based on these two diagnostics that are applicable to the galaxies in our sample, we cannot define a clear subsample of narrow AGN. For this reason, we keep these sources in the final sample analyzed in this paper. 

\begin{figure}[t!]
    \centering
\includegraphics[width=\columnwidth]{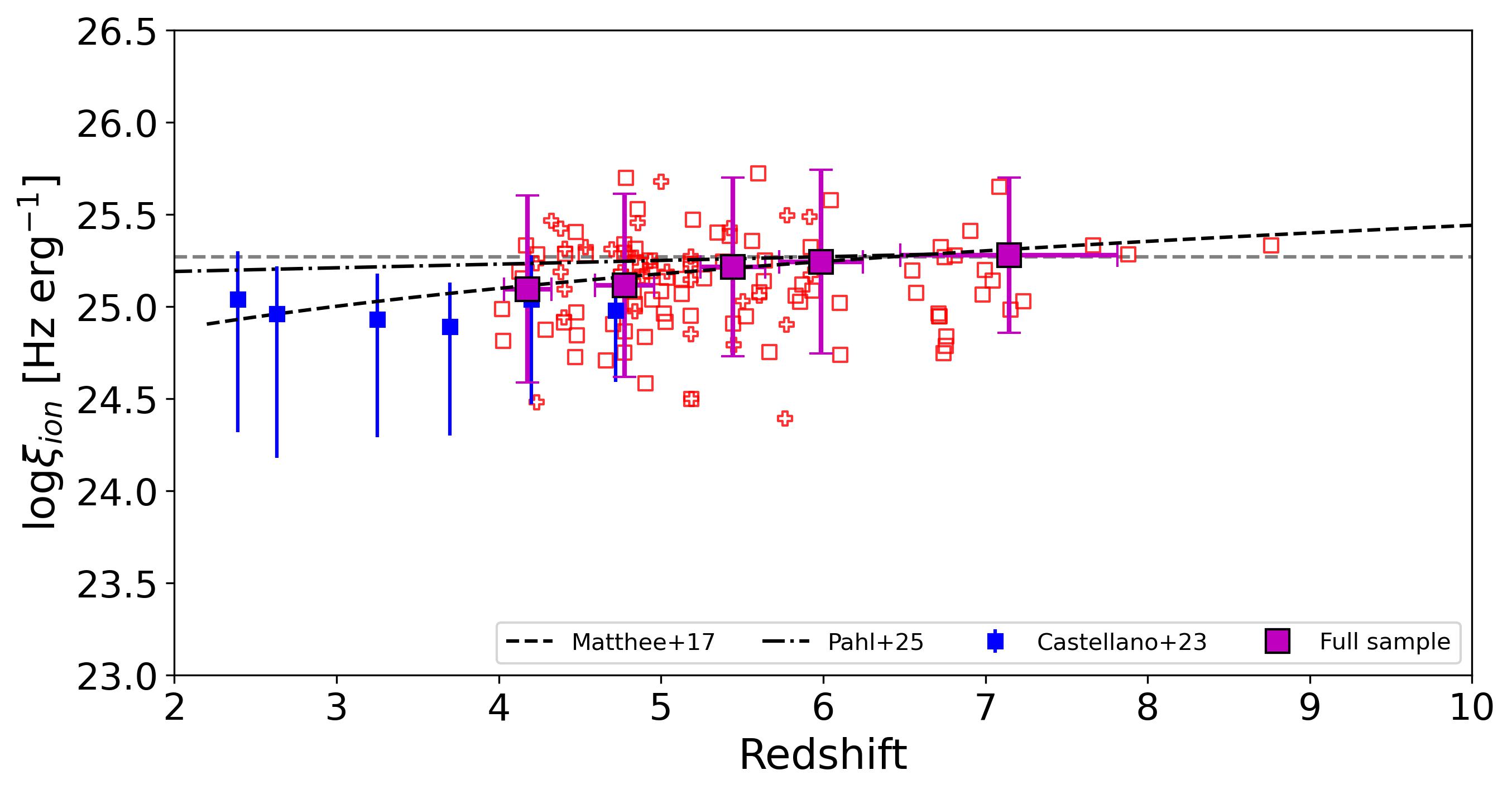}
    \caption{AGN candidates in the evolution of \xiion with redshift. The symbols are the same as in Fig. \ref{fig:xi_z}. The red symbols are the AGN candidates from Fig. \ref{fig:OHNO}.}
    \label{fig:xi_z_agn}
\end{figure}

\begin{figure}[t!]
    \centering
    \includegraphics[width=\columnwidth]{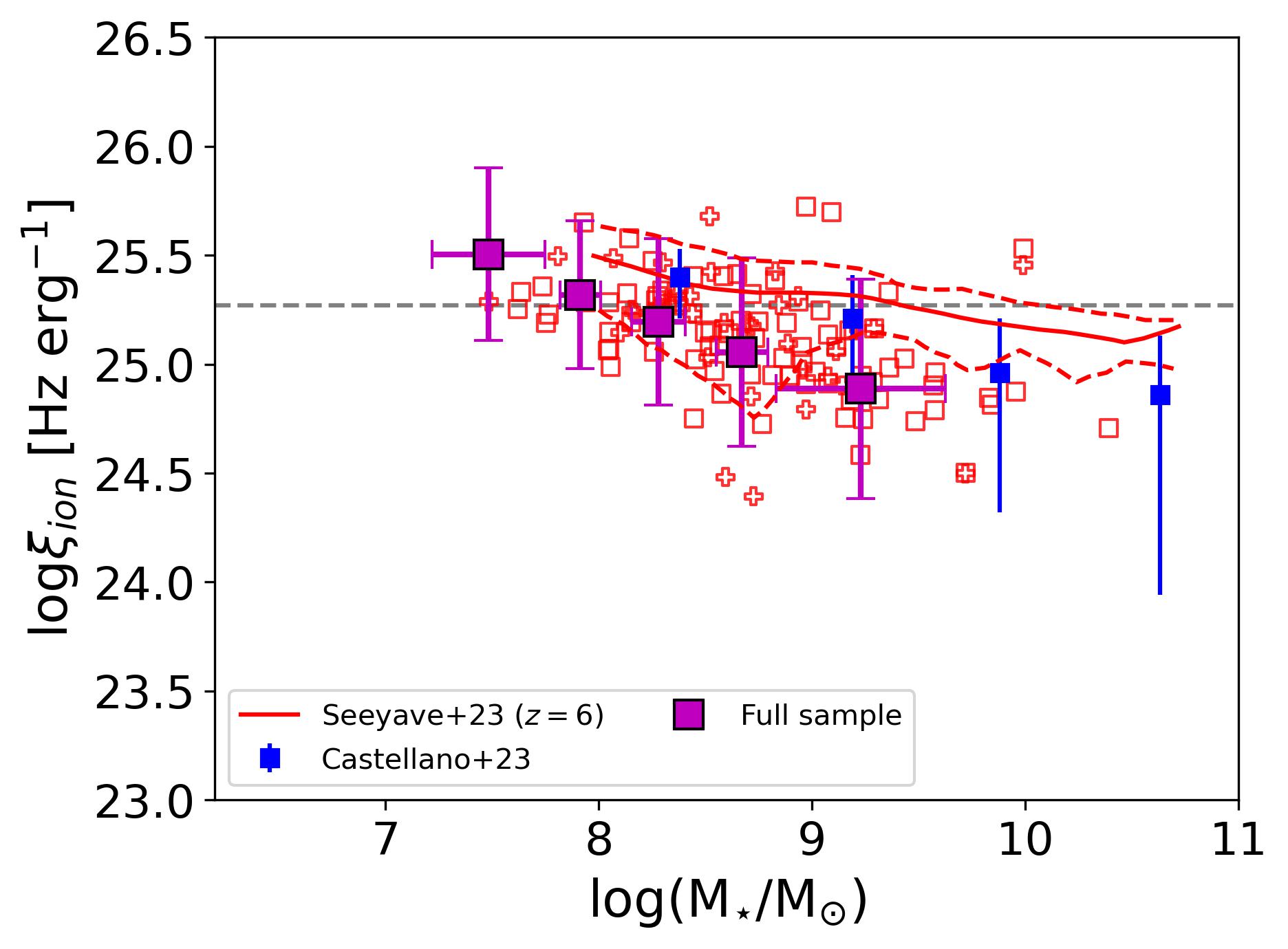}
    \caption{AGN candidates in the relation of \xiion with stellar mass. The symbols are the same as in Fig. \ref{fig:xi_mass}. The red symbols are the AGN candidates from Fig. \ref{fig:OHNO}.}
    \label{fig:xi_mass_agn}
\end{figure}

\begin{figure}[t!]
    \centering
    \includegraphics[width=\columnwidth]{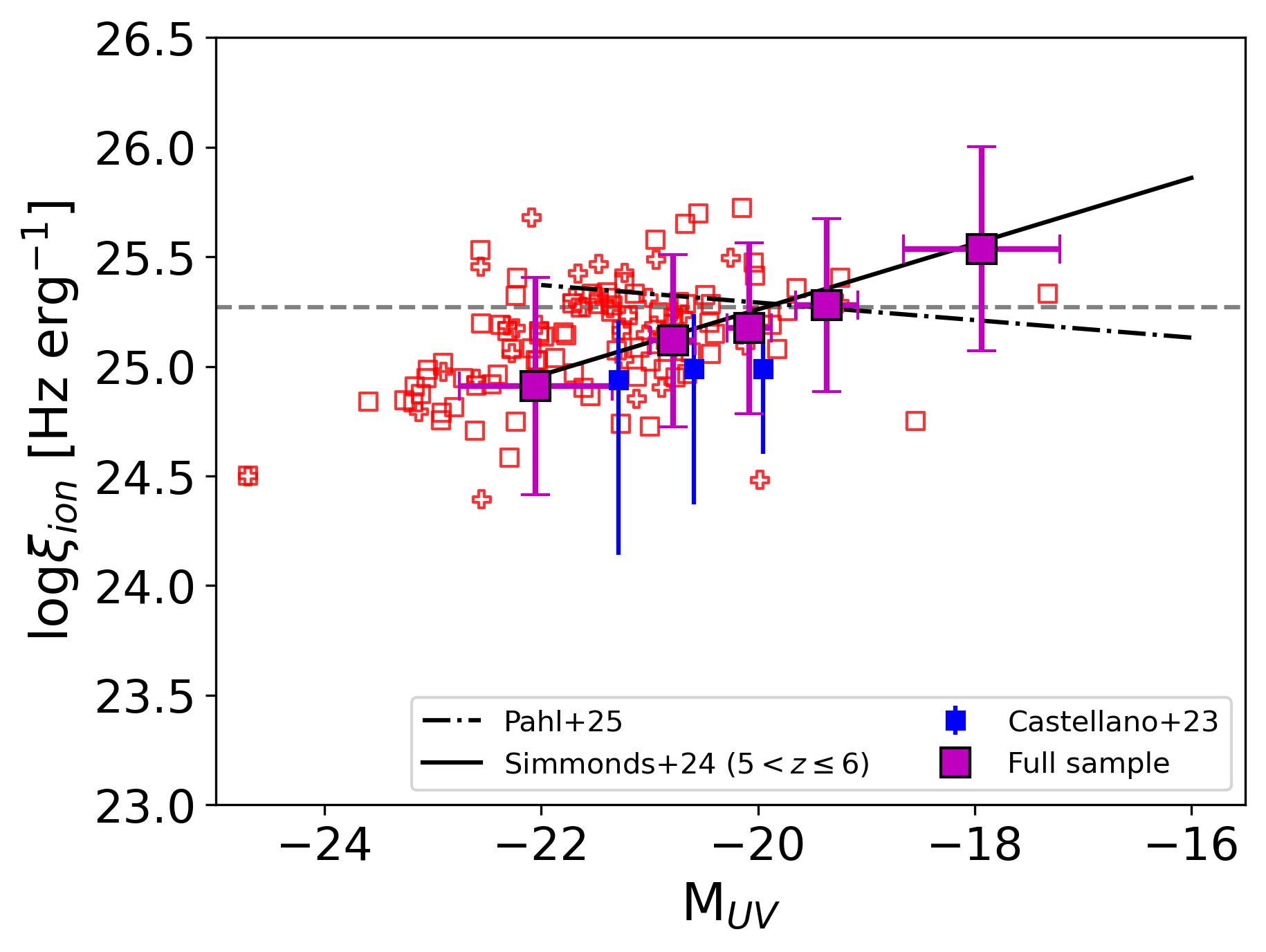}
    \caption{AGN candidates in the relation of \xiion with M$_{UV}$. The symbols are the same as in Fig. \ref{fig:xi_MUV}. The red symbols are the AGN candidates from Fig. \ref{fig:OHNO}.}
    \label{fig:xi_muv_agn}
\end{figure}

We explore if there are biases in the relation we found is \ref{sec:results_all} due the inclusion of these AGN candidates in our sample. In Fig. \ref{fig:xi_z_agn}, we show the location of the AGN candidates based on the OHNO diagram, in the relation of \xiion with redshift, which was shown in Fig. \ref{fig:xi_z}. We see that they show a wide range of redshifts from $z\sim 4-8$ and a wide range of \xiion values. They follow the trend of the median values of the full sample, which suggests there is no bias due to these candidates in the overall trend. Similar results are found if we analyze the relation of \xiion with the stellar mass which is shown in Fig. \ref{fig:xi_mass_agn}. We show they have a wide range of stellar masses $\gtrsim 10^{7.5}$\Msun\, and \xiion values. Similarly, they follow the trend that is observed with the median values of the full sample. Finally, in Fig. \ref{fig:xi_muv_agn} we check the location of the AGN candidates in the relation of \xiion with M$_{UV}$. Similarly, we find that they show a wide range of M$_{UV}$ and \xiion values. They are in general bright (M$_{UV}\lesssim$-19) compared with the full sample which suggests this may be the reason [NeIII]$\lambda$3870 is detected in these candidates rather than being ionized by AGN. We also find that these candidates follow the trend observed with the median values of the full sample. Overall, we find that including these not secure AGN candidates in our sample does not affect the trends we find in this paper. 

\section{\xiion from Balmer lines}\label{appendix:ha_hb}
\begin{figure}[t!]
    \centering
\includegraphics[width=\columnwidth]{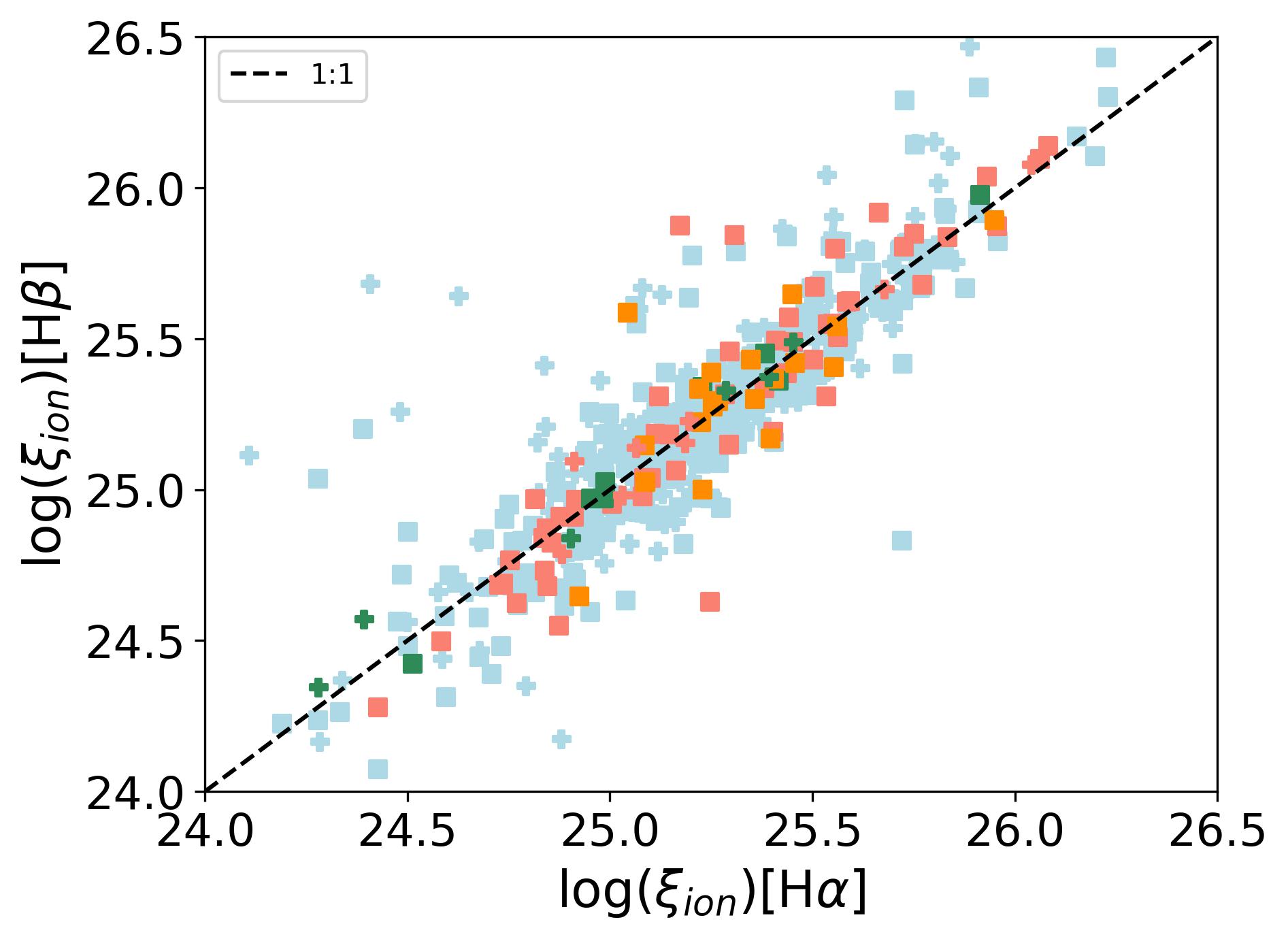}
    \caption{Comparison between \xiion determined from \halpha and \hbeta luminosities. The symbols are individual galaxies in the sample with simultaneous detection of \halpha and \hbeta. The symbols are the same as in Fig. \ref{fig:xi_z}.}
    \label{fig:xi_ha_hb}
\end{figure}
In Fig. \ref{fig:xi_ha_hb} we show a comparison between the values of \xiion determined from \halpha and \hbeta for the subsample of galaxies in the full sample with simultaneous detection of both Balmer lines. We find that there is a good agreement between both estimations with a median difference of 0.05 dex. Due to this fact, we are not introducing a bias when we mix both estimations in the galaxies where \halpha is not detected.

\section{Observed Balmer fluxes}\label{appendix:balmer_flux}

\begin{figure*}
    \centering
    \includegraphics[width=0.8\linewidth]{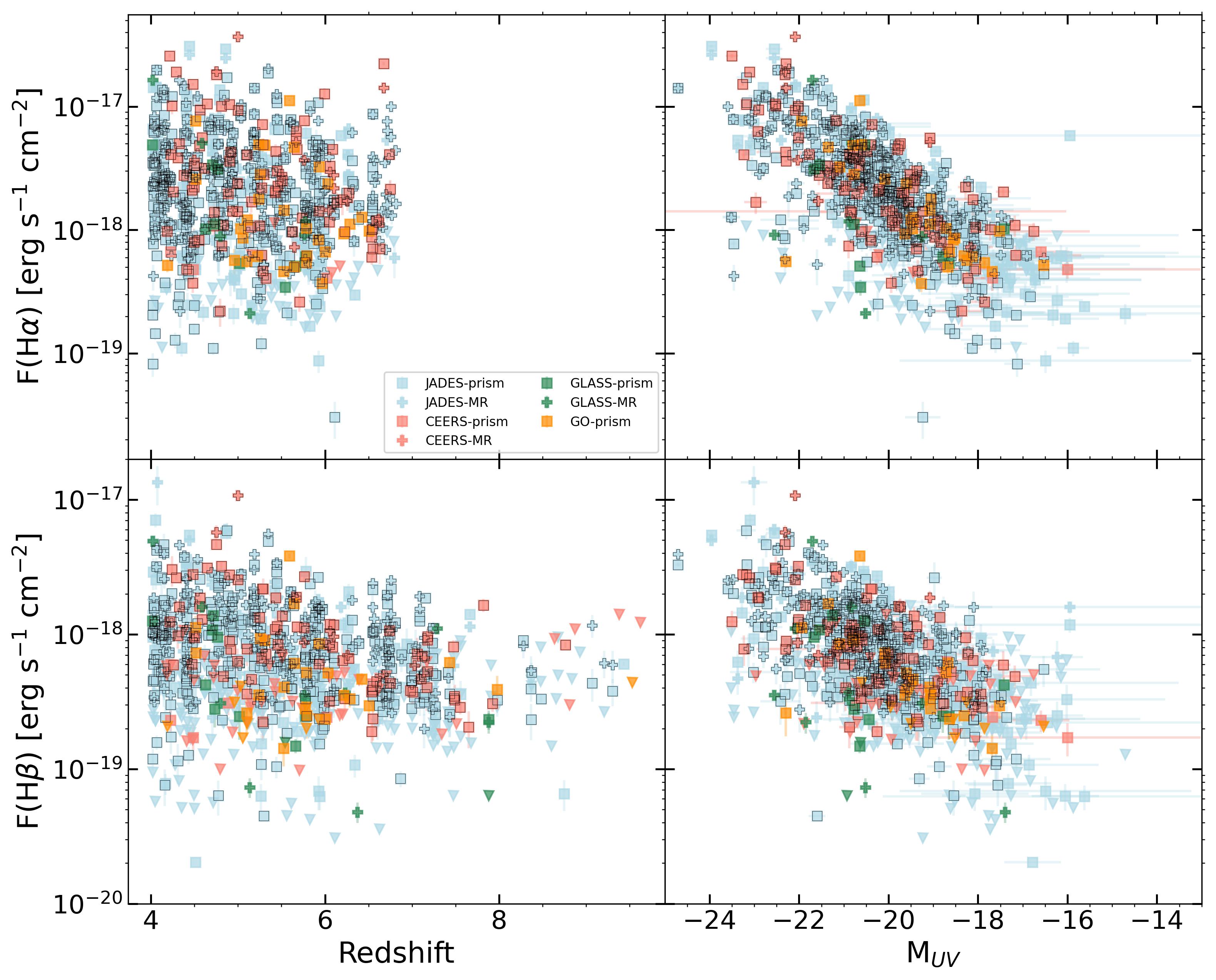}
    \caption{{Observed fluxes of \halpha (top panels) and \hbeta (bottom panels) as a function of redshift (left panels) and M$_{UV}$ (right panels). The symbols are the same as in Fig. \ref{fig:xi_z}.}}
    \label{fig:balmer_fluxes}
\end{figure*}

{In Fig. \ref{fig:balmer_fluxes} we show the relation between the observed fluxes of Balmer lines as a function of redshift and M$_{UV}$. It can be noticed that both \halpha and \hbeta cover a wide range of fluxes ($\sim2 $dex). For \halpha this distribution is uniform across all the redshifts ($z\lesssim7$) while for \hbeta we measured fluxes $<2\times10^{-18}$ erg s$^{-1}$ cm$^{-2}$ for galaxies at $z\gtrsim7$ and some of them are upper limits in flux. We also note that most of the complete subsample includes galaxies up to $z\sim7$ for both lines.}

{Regarding the relation with M$_{UV}$ we find a decreasing trend of the fluxes with increasing M$_{UV}$ which indicates that the faintest UV galaxies show also fainter Balmer lines compared to brighter UV galaxies. This is in agreement with the distribution along the MS which is shown in Fig. \ref{fig:MS}. In our sample we do not include UV bright passive galaxies as can be seen in Fig. \ref{fig:balmer_fluxes}. We also note that most of the galaxies in the complete subsample shows M$_{UV}\lesssim-18$. We need to take into account these ranges in redshift and M$_{UV}$ when analyzing the results since in these ranges we have a complete sample and the results are more robust.}

\end{appendix}
\end{document}